\documentclass[aps,preprint,showpacs,amssymb,superscriptaddress]{revtex4}
\usepackage{graphicx}
\usepackage{bm}
\usepackage{bm}
\begin{document}
\draft
\title{Critical Casimir Effect in superfluid wetting films}
\author{A.~Macio\l ek}
\affiliation{Max-Planck-Institut f{\"u}r Metallforschung, Heisenbergstr.~3, D-70569 Stuttgart, Germany}
\affiliation{Institut f{\"u}r Theoretische und Angewandte Physik,
Universit{\"a}t Stuttgart, Pfaffenwaldring 57, D-70569 Stuttgart, Germany}
\affiliation{Institute of Physical Chemistry,
             Polish Academy of Sciences, Kasprzaka 44/52,
            PL-01-224 Warsaw, Poland}
\author{A.~Gambassi}
\affiliation{Max-Planck-Institut f{\"u}r Metallforschung, Heisenbergstr.~3, D-70569 Stuttgart, Germany}
\affiliation{Institut f{\"u}r Theoretische und Angewandte Physik,
Universit{\"a}t Stuttgart, Pfaffenwaldring 57, D-70569 Stuttgart, Germany}
\author{S.~Dietrich}
\affiliation{Max-Planck-Institut f{\"u}r Metallforschung, Heisenbergstr.~3, D-70569 Stuttgart, Germany}
\affiliation{Institut f{\"u}r Theoretische und Angewandte Physik,
Universit{\"a}t Stuttgart, Pfaffenwaldring 57, D-70569 Stuttgart, Germany}
\date{\today}

\begin{abstract}
Recent experimental data for the complete  wetting behavior of pure $^4$He 
and of $^3$He-$^4$He mixtures exposed to solid substrates show 
that there is a change of the corresponding  film thicknesses $L$ 
upon approaching thermodynamically  the  $\lambda$-transition and 
 the tricritical end  point, respectively,  which can 
be attributed to critical
 Casimir forces $f_C$.
 We calculate the scaling functions $\vartheta$ of $f_C$
 within models representing the  corresponding universality classes.
 For the  mixtures our
analysis  provides an understanding of the  rich behavior of $\vartheta$ 
deduced from  the experimental data and predicts  the crossover behavior
between the tricritical point and the $\lambda$-transition of 
 pure $^4$He which are connected by a line of critical points.
The formation of a  'soft-mode' phase within the wetting  films gives rise to 
a  pronounced maximum of $f_C$ below the tricritical point as 
observed experimentally.
Near the tricritical point we find logarithmic corrections 
 $\sim L^{-3}(\ln L)^{1/2}$ for the leading behavior of $\vartheta$ 
dominating  the 
 contributions from the background dispersion forces.  
\end{abstract}
\maketitle

\section{Introduction}
\label{sec:int}
 
There is  growing experimental evidence for the analogue 
of the electromagnetic Casimir effect \cite{casimir} in various critical
condensed matter  systems 
\cite{garcia:99:0,law:99:0,garcia:02:0,balibar:02:0,pershan,garcia:06:0}.
 In wetting experiments the confinement of critical fluctuations within 
 an adsorbed liquid film gives rise to an effective Casimir
 force $f_C$ between the substrate-liquid and the liquid-vapor interfaces
of the liquid film \cite{fdg,krech:99:0,krech:91}.
Near the critical end point of the liquid the emerging Casimir force adds 
to the omnipresent dispersion forces and thus leads to a change of the 
thickness of the complete wetting film. From this response one can infer 
the Casimir force by subtracting the effect of the background forces
 which varies smoothly near the critical end point with  temperature
$T_c$. 
In accordance with  
 finite-size scaling theory \cite{privman} 
this force $f_C$  per unit area and in units of $k_BT_c$ can be expressed
 in terms of a  universal scaling function $\vartheta$; its  shape depends
 sensitively on the type of boundary conditions (BC) \cite{krech:99:0}
 and thus on the surface universality classes the confining surfaces
 belong to \cite{diehl:86:0}. 

Capacitance measurements of the equilibrium 
thickness of $^4$He wetting films near the superfluid temperature 
$T_{\lambda}$ of the critical end point of the $\lambda$-line 
\cite{garcia:99:0,garcia:06:0} quantitatively  support
the theoretical predictions of  $f_C$ for  the bulk   universality 
class of the XY model with  {\it symmetric} Dirichlet-Dirichlet BC 
 $(O,O)$ forming the so-called ordinary (O) surface universality class
\cite{diehl:86:0}.
Such  BC correspond to  the case that  the  quantum-mechanical
 wave function of  the superfluid state vanishes at 
 both interfaces, giving rise to an attractive 
 Casimir  force  $(f_C<0)$  \cite{krech:99:0,krech:91}.
However, the available  theoretical results  have a  limited range
 of applicability, i.e.,   $T\ge T_{\lambda}$ and  $T\ll T_{\lambda}$.
Above and at $T_{\lambda}$ explicit field-theoretical calculations within 
the $\epsilon$-expansion scheme are  available \cite{krech:92:a,krech:92:b}. 
For temperatures well below $T_{\lambda}$ there are calculations 
which  take into account capillary-wavelike surface fluctuations  
in the asymptotic limit 
of thick  films, predicting a levelling off  of the scaling
 function for large negative scaling variables \cite{kardar:04}, i.e., $T\ll T_{\lambda}$, in qualitative agreement with the experimental observations.
So far there are no theoretical  results available for the critical
 region below $T_{\lambda}$ which provide an  understanding
 of the  deep minimum of the experimental scaling function (ca. 20 times deeper
than its  value at $T_{\lambda}$).

$^3$He-$^4$He mixtures near their tricritical end  point
 (see Fig.~12 in Ref.~\cite{krech:92:b}) are another
 critical system for   which %
wetting experiments have been 
 performed recently \cite{garcia:02:0,balibar:02:0}. 
The tricritical end  point with temperature $T_t$ is %
the point 
in the $^3$He-$^4$He  phase diagram where the line
signalling the onset of superfluidity 
joins the top of the two-phase  coexistence region for phase separation
into a $^4$He-rich superfluid phase and a $^3$He-rich normal phase. 
The mixture
belongs  to a  bulk universality class different from that 
one of pure $^4$He and,   because its  upper critical spatial dimension 
 $d^{\ast}$ equals 3, the actual physical system  is characterized 
by rational mean-field critical exponents (up to logarithmic corrections)
\cite{riedel:72:0,LawSar}.
The capacitance  measurements
 of the wetting  film thickness 
of the mixture
reveal 
a  {\it repulsive} Casimir force $f_C$ around  the tricritical end  point
which  suggests {\it non-symmetric} BC for
 the superfluid order parameter (OP). 
The probable  physical mechanism behind  such a BC
is  that within   $^3$He-$^4$He  wetting films
a $^4$He-rich layer forms near the substrate-liquid
 interface, which may become superfluid already above
 the line of onset of superfluidity in the bulk ~\cite{laheurte:78:0} whereas the lighter $^3$He has
a preference for the liquid-vapor interface.
Thus the two interfaces impose a nontrivial concentration profile
which in turn couples to the superfluid OP. 

For this system, recently  \cite{maciolek:06:0}  we briefly reported 
 explicit model calculations which demonstrate 
 that the concentration
 profile indeed induces indirectly {\it non-symmetric} 
 BC for the superfluid OP.
For  symmetry-breaking (+) BC  at the substrate-liquid  interface
and  Dirichlet $(O)$ BC at the liquid-vapor interface 
 we calculated the Casimir force and found a semiquantitative agreement with
the  experimental data given in Ref.~\cite{garcia:02:0}.
 Moreover, we formulated theoretical predictions
for the behavior of $f_C$ in the crossover regime 
 between the tricritical point 
and the $\lambda$-transition of pure $^4$He which are connected
by a line of  critical points and provided  the universal leading behavior
of the Casimir force at the tricritical point.

The purpose of the present study  is to elucidate  the details of 
the two complementary approaches  used in Ref.~\cite{maciolek:06:0}
 and to extend them in order to obtain new results 
both for the tricritical  $^3$He-$^4$He mixture and the critical pure
$^4$He.
The
presentation is organized %
as follows:
In Sec.~\ref{sec:2} we discuss the universal properties of the
Casimir force.
As already mentioned above, for the present tricritical behavior the
upper critical dimension $d^{\ast}$ equals 3 and therefore 
the thermodynamic functions of three-dimensional  systems exhibit 
 power-law behaviors with 
critical exponents taking their  classical values.
However,  logarithmic corrections to the mean-field  (MF)
behavior are expected  under experimental conditions \cite{LawSar}.
Using field-theoretical methods and renormalization-group (RG)
analyses we obtain  the  leading asymptotic behavior
of the Casimir force at the tricritical point. As a function of the
 film thickness  $L$ it has the  form of a 
power law multiplied by a fractional power of a logarithm and by the
{\it universal} Casimir amplitude.
In addition, we also derive  the form of the
finite-size scaling 
  for the Casimir force in the vicinity of the tricritical point.
 As expected
\cite{LawSar},  also the arguments of the 
associate
scaling function acquire
 logarithmic corrections. These scaling functions are compared with the ones
deduced from the experimental data in Ref.~\cite{garcia:02:0}.
In  Sec.~\ref{sec:3}  we study within mean-field theory (MFT) films of the lattice vectoralized Blume-Emery-Griffiths (VBEG) model ~\cite{maciolek:04:0} which
belongs to
the same  universality class as the 
$^3$He-$^4$He mixture but  is
simple enough to allow  
for systematic studies
of $f_C$ along all 
 thermodynamic paths followed in the wetting
 experiments of Ref.~\cite{garcia:02:0}. This facilitates the exploration
of the crossover
 between the tricritical point $T_t$ and
 the line of critical points and the coexistence region below
$T_t$. This enables us to follow the Casimir force upon  continuously
switching  the bulk universality class  (from tricritical to critical)
 by changing the concentration  of the $^3$He-$^4$He mixture. 
 The scaling functions corresponding to
 thermodynamic paths of constant concentration  of the two components of
the   $^3$He-$^4$He mixtures are calculated and %
compared 
with the corresponding experimental data in Ref.~\cite{garcia:02:0}.  As a limiting case the 
 VBEG model can describe also  
a film of 
pure $^4$He  which  is studied 
in Sec.~\ref{sec:4} within MFT. 
The scaling 
function of  the 
corresponding
Casimir force 
is obtained in  the critical
 region below $T_{\lambda}$  and compared with that 
one
extracted 
from  the experimental data in  Ref.~\cite{garcia:99:0}. 
We also 
compare these results with the mean-field predictions which follow
from the Landau-Ginzburg theory in the  film geometry with suitable BC.
In Sec.~\ref{sec:5}
we discuss  
the theoretical results obtained within the VBEG model
and assess 
their relevance
for interpreting the experimental data. 
We conclude 
with a
summary and an outlook in Sec.~\ref{sec:6}.

\section{Universal properties}
\label{sec:2}
For film geometries, in this section we investigate  
the universal properties of the Casimir force near tricriticality.
 In general  two-component systems are characterized by the ordering
 density ${\bf \Phi}$ and its conjugate field $h$, 
and by a non-ordering density
$x$ and its conjugate field $\Delta$.
 For liquid $^3$He-$^4$He mixtures, 
${\bf \Phi}$, $x$, and $\Delta$  correspond to the superfluid OP, 
 to the $^3$He concentration  and
to the difference between the chemical potentials of the $^3$He and $^4$He components, respectively,  whereas the field $h$ conjugate to the superfluid OP is experimentally not accessible.

\subsection{Scaling function from  Landau-Ginzburg theory}
\label{subsec:12}
In order to capture universal properties  we consider  the  
 standard dimensionless  $O(n)$-symmetric 
 Landau-Ginzburg (LG)
Hamiltonian  for a tricritical  system
 in the  film geometry:
\begin{equation}
\label{eq:1}
{\cal H}[{\bf \Phi}]=\int {\rm d}^{d-1}x\int_0^L{\rm d}z\left\{ \frac{1}{2}(\nabla {\bf \Phi})^2+\frac{r_0}{2}{\bf\Phi}^2+\frac{u_0}{4!}({\bf\Phi}^2)^2+\frac{v_0}{6!}({\bf\Phi}^2)^3 \right\},
\end{equation}
where $L$ is the film thickness, ${\bf \Phi}$ is the $n$-component 
order parameter  OP ($n=2$ corresponds to the XY universality class),
and $z$ is the coordinate normal to  the confining surfaces;
 $r_0$,  $u_0$, and 
 $v_0$ are  bare coupling  constants
 depending, {\it inter alia}, 
on the temperature $T$ and  the non-ordering field $\Delta$.
 $r_0(u_0)=0$ and $u_0>0$ define the critical line, whereas at the tricritical point one has $r_0=u_0=0, v_0>0$.
The semi-infinite version of Eq.~(\ref{eq:1}) 
has been studied in the context of  surface critical behavior
\cite{eisen:88}. 
In the  film geometry the Casimir force per area $A$ of the cross section of the film and in units of $k_BT_t$,
\begin{equation}
\label{eq:1a}
f_C\equiv -(\partial f^{ex}/\partial L)=\langle {\cal T}_{zz}\rangle,
\end{equation}
 is given by the thermal average of the  stress tensor
component ${\cal T}_{zz}$~\cite{krech:99:0}:
\begin{equation}
\label{eq:1a1}
 f^{ex}(L)\equiv (f-f_b)L/(k_BT_t)
\end{equation}
where $f$ 
is the total free energy of the film per volume $V=LA$ %
and  $f_b$ is the bulk free energy density.
For large $L$ the excess free energy can be decomposed into surface and
 finite-size contributions:  $f^{ex}(L)=f_{s,1}+f_{s,2}+\delta f(L)$.
The stress tensor is given by~\cite{krech:99:0}
\begin{equation}
\label{eq:1b}
{\cal T}_{ij}=\partial_i{\bf \Phi}\cdot \partial_j{\bf \Phi}-\delta_{ij}{\cal L}-(d-2)/(4(d-1))(\partial_i\partial_j-\delta_{ij}\nabla^2){\bf \Phi}^2,
\end{equation}
where ${\cal L}$ is the  integrand in Eq.~(\ref{eq:1}).
In what follows we assume ${\bf \Phi}=(m(z),0,\ldots,0)$, i.e., we neglect helicity.
For non-symmetric  BC  its relevance for the behavior 
of the Casimir force is not clear because the OP has the additional freedom
to rotate across the film by a position dependent angle $\phi(z)$; the analyses
of the role of  helicity is left  for  future research. 
Within MFT for the LG  Hamiltonian, the determination 
 of the tricritical Casimir force  in the  film geometry starts
from the Euler-Lagrange equation
\begin{equation}\label{ELtri}
m''(z) = r_0 m(z) + \frac{u_0}{6}m^3(z) + \frac{v_0}{120}m^5(z).
\end{equation}
As discussed in Sec.~\ref{sec:int},  $(+,O)$ boundary conditions,
 with the substrate at $z<0$ and vapor at $z>L$,
\begin{equation}\label{BC34}
m(0) = +\infty \quad \mbox{and} \quad m(L) = 0
\end{equation}
 are supposed to mimic the experimental system of $^3$He-$^4$He wetting films
as studied in  Ref.~\cite{garcia:02:0}.
According to Eq.~(\ref{eq:1b}) the stress tensor component ${\cal T}_{zz}$ evaluated 
within   MFT and with  ${\bf \Phi}=(m(z),0,\ldots,0)$ for the
OP  (in the present MF approach  we omit the brackets
 $\langle\cdot \rangle$  indicating the thermal average) yields
\begin{equation}\label{force}
{\cal T}_{zz} = \frac{1}{2}(m'(L))^2.
\end{equation}
In deriving this expression
we have used the property that 
 ${\cal T}_{zz}=const$ throughout the film including the surfaces
 and we have chosen $z_0=L$ as the point of reference at which ${\cal T}_{zz}$ is evaluated. 
Accordingly,  the first integral of Eq.~(\ref{ELtri}) is  given by
\begin{equation}\label{ELtri1}
(m_{+,O}'(z))^2 = 2 {\cal T}_{zz} + r_0 m_{+,O}^2(z) + \frac{u_0}{12}m_{+,O}^4(z) + \frac{v_0}{360}m_{+,O}^6(z).
\end{equation}
Dimensional analysis yields that, at the upper critical dimension
 $d=d^{\ast}=3$, $ m(z,L,r_0,u_0,v_0)$ can be expressed in terms of  a dimensionless  scaling function $\varphi_{+,O}$:
\begin{equation}\label{scaltri}
m_{+,O}(z,L,r_0,u_0,v_0)= \left(\frac{v_0}{360}\right)^{-1/4} L^{-1/2} \varphi_{+,O}(z/L,r_0L^2,u_0L;v_0) ,
\end{equation}
where $v_0$ is dimensionless. Similarly, within this approach the normalized Casimir force can be expressed in terms of a dimensionless scaling function $\vartheta_{+,O}$:
\begin{equation}
\label{scaltriC}
{\cal T}_{zz}=f_C(L,r_0,u_0,v_0)=\left(\frac{v_0}{90}\right)^{-1/2}L^{-3} \vartheta_{+,O}^{MF}(r_0L^2,u_0L,v_0).
\end{equation}
 Equation~(\ref{ELtri1}) can be written in terms of these  scaling functions
$\varphi_{+,O}$ and $\vartheta_{+,O}$:
\begin{equation}\label{ELtri1scal}
(\varphi_{+,O}'(x))^2 = \vartheta_{+,O}^{MF} + r_0L^2\ \varphi_{+,O}^2(x) +
\left(\frac{5}{2v_0}\right)^{1/2} u_0L\ \varphi_{+,O}^4(x) + \varphi_{+,O}^6(x),
\end{equation}
where $x = z/L$. In turn, Eq.~(\ref{ELtri1scal}) can be integrated directly
yielding
the implicit equation
\begin{equation}\label{implicit}
1 = \int_0^{\infty}\frac{{\rm d}\varphi}{\sqrt{\vartheta_{+,O}^{MF} + r_0L^2 \varphi^2 +
\left(\frac{5}{2v_0}\right)^{1/2} u_0L \varphi^4 + \varphi^6}}
\end{equation}
for the scaling function $\vartheta_{+,O}^{MF}(r_0L^2,u_0L,v_0)$. Note that the coupling
constant  $v_0 > 0$
remains undetermined within mean-field theory and enters into 
$\vartheta_{+,O}^{MF}$ only in the combination $v_0^{-1/2}u_0L$. Under the renormalization
group flow,  at the
upper critical dimension ($d^{\ast}=3$) the renormalized coupling
constant $v$ 
associated with $v_0$
tends 
to its fixed point value
$v^* = 0$. This RG  flow generates
logarithmic corrections to scaling due to the singular dependence of the
scaled quantities on $v$
 (see, e.g.,  Eqs.~(\ref{scaltri}) and (\ref{scaltriC})).
With the transformation
\begin{equation}\label{transform}
\varphi = \left(\vartheta^{MF}_{+,O}\right)^{1/6} p
\end{equation}
for the integration variable one can rewrite Eq.~(\ref{implicit}) in the
 more convenient but still implicit form
\begin{equation}\label{implicit1}
\left(\vartheta_{+,O}^{MF}\right)^{1/3} = \int_0^{\infty}\frac{{\rm d}p}{\sqrt{1 + a p^2 + b p^4 + p^6}},
\end{equation}
where the dimensionless  parameters $a$ and $b$ are given by
\begin{equation}\label{ab}
a = r_0L^2 \left(\vartheta_{+,O}^{MF}\right)^{-2/3} \quad\mbox{and}\quad
b = \left(\frac{5}{2v_0}\right)^{1/2} u_0L \left(\vartheta_{+,O}^{MF}\right)^{-1/3}.
\end{equation}
The numerical evaluation of the scaling function 
 amounts to the following steps:
(1) specifying  values for $a$ and $b$, (2) evaluating
 $\vartheta_{+,O}^{MF}$ from 
Eq.~(\ref{implicit1}), (3) determining  the values of the two scaling variables
$r_0L^2$ and $v_0^{-1/2}u_0L$ from Eq.~(\ref{ab}).

From the symmetry properties
of the order-parameter profile for the symmetry breaking opposing
 boundary conditions $(+,-)$ it is obvious that within MFT the force 
for a film of thickness $L$ in this case
can be obtained from Eqs.~(\ref{implicit1}), holding for $(+,O)$ BC,
 and (\ref{scaltriC}) by replacing  $L \mapsto L/2$ therein.   This implies
$\vartheta^{MF}_{+,-}(x,y)=8\vartheta^{MF}_{+,O}(x/4,y/2)$. 
In the following we shall refer only to the $(+,O)$ BC and drop the corresponding index.

 The precise  dependence of $r_0$ and $u_0$ on the 
thermodynamic fields $T$ and $\Delta$ is not known. Therefore
 it is not obvious how to follow in terms of these variables a
 specified path in the phase diagram such as 
the experimental path of fixed $^3$He concentration. However, 
assuming that $r_0$ and $u_0$ are analytic functions of $T$ 
and $\Delta$ in the neighborhood of the phase transition one can use
 the expansion
 \cite{LawSar}:
\begin{equation}\label{expan}
r_0 = A(\Delta)(T-T_{\lambda}(\Delta))+O((T-T_{\lambda}(\Delta))^2) \quad\mbox{and}\quad
u_0 = B(\Delta)+O((T-T_{\lambda}(\Delta))),
\end{equation}
where $T_{\lambda}(\Delta) $ denotes  the critical temperatures of the line of continuous phase transitions  as a function of $\Delta$, and  $B(\Delta)$ and $ A(\Delta)$
are  positive and non-zero on this  line;  $B(\Delta)=0$
 at the tricritical point.

In view of comparisons with experimental data, which we shall discuss later,
 it is useful
 to mention the relation between the  parameters $r_0$ and $u_0$ and the
experimentally controllable  thermodynamic fields 
$T-T_t$ 
and $\Delta-\Delta_t$ where $\Delta=\Delta_t$ at the tricritical point 
and $T_{\lambda}(\Delta_t)=T_t$.  These ``deviating fields'' are not the proper scaling
fields and 
it was shown \cite{riedel:72} that  a suitable (dimensionless) choice
 is provided by
\begin{equation}\label{scalfields} 
 t\equiv (T-T_t)/T_t  \qquad\mbox{and}\qquad  g\equiv (\Delta-\Delta_t)/(k_BT_t)+a't,
\end{equation}
where $a'$ is the slope of the line tangential to the phase boundary at the tricritical point.
Thus for  $t\to 0$ with  $g=0$ the tricritical point 
 is approached tangentially to
 the phase boundary. Instead of $t$ one could also  use a scaling variable which is 
orthogonal to the loci $g=0$;   this would  not  affect the leading
singular behavior for  $t, g \to 0$ \cite{LawSar}. 
Near the tricritical point  $B(\Delta)$,  $ A(\Delta)$,  and 
$T_{\lambda}(\Delta)$  
can be expanded  in terms of $g$ and $t$. Using Eq.~(\ref{scalfields}) 
one has  $T-T_{\lambda}(\Delta)=T-T_t+(a'k_B)^{-1}(\Delta-\Delta_t)+O((\Delta-\Delta_t)^2)=(a')^{-1}T_tg+O((\Delta-\Delta_t)^2)$. Expressing
$\Delta $ and $T$ as a function of $t$ and $g$ %
one finds:
\begin{equation}\label{expan1}
r_0 = A_1g+A_2t^2+O(g^2,gt) \quad\mbox{and}\quad
u_0 = B_1t+B_2g+O(gt,g^2,t^2)
\end{equation}
where $A_1>0, B_1>0, A_2,$ and $B_2$ are constants. 
Due to the analytic structure of Eq.~(\ref{expan}) and because
$(\Delta-\Delta_t)=k_BT_t(g-a't)$ the coefficient $r_0$ does not
 contain a  term  linear in $t$  so that 
    $u_0\sim t+O(t^2)$ if  $r_0=0$. On the other hand $r_0\sim g+O(g^2)$ if $u_0=0$.
 
\subsection{Logarithmic corrections at $T=T_t$}
\label{subsec:22}

At the tricritical point $a=b=0$
Eq.~(\ref{implicit1}) reduces to
\begin{equation}\label{implicit2}
(\vartheta^{MF})^{1/3} =\displaystyle \int_0^{\infty}{\rm d}p/\sqrt{1 + p^6}\simeq 1.40218.
\end{equation}
Accordingly,
in units of $Ak_BT_t$,  the MFT  result for the tricritical 
Casimir force $f_C^t$ in the case of $(+,O)$ BC is (see Eq.~(\ref{scaltriC}))
\begin{equation}
\label{eq:2a_1}
f_{C,t}^{MF}\simeq 2.75684\left( 90/v_0\right)^{1/2}L^{-3}.
\end{equation}
In  $d=3-\epsilon $ the  MFT result at tricriticality (Eq.~(\ref{eq:2a_1}))
 yields
 the leading contribution in a perturbation series, i.e.,
\begin{equation}
\label{eq:pertser}
\langle {\cal T}_{zz}\rangle =\langle {\cal T}_{zz}\rangle _0+\langle {\cal T}_{zz}\rangle _1+O(v_0^{1/2})=
\left(\frac{v_0}{90}\right)^{-1/2}t_{zz}+\langle {\cal T}_{zz}\rangle _1+O(v_0^{1/2})
\end{equation}
where both  $t_{zz}\equiv  2.75684 L^{-3}$ and  $\langle{\cal  T}_{zz}\rangle _1$
do not depend on $v_0$. After removing ultraviolet  singularities 
  via renormalization (R) the  asymptotic scaling behavior of $f_C$  follows from 
substituting  the renormalized 
$v$ by the  appropriate  fixed-point  value
  $v^{\ast}\propto \epsilon$.
At $d=d^{\ast}$, and under spatial rescaling by a dimensionless factor  $\ell$,
$v$ flows to its RG fixed point value
$v^* = 0 $ according to  \cite{eisen:88}
\begin{equation}
\label{eq:2a_2}
\bar{v}(\ell)=\frac{240 \pi^2}{3n+22}\left[\frac{1}{|\ln \ell|}+c\frac{\ln|\ln\ell|}{\ln^2\ell}+\ldots\right],
\end{equation}
where $\bar{v}(\ell)$ is the running coupling constant
 with the initial condition
$\bar{v}_R(\ell=1)=v_R$.
With the  rescaling factor $\ell =l_0/L$, where $l_0$ is a
 microscopic length 
scale of the order of a  few \AA, ~this yields a logarithmic correction
 to the power-law dependence on $L$ of  the tricritical 
Casimir force:
\begin{equation}
\label{eq:as}
f_C^t\simeq 0.54(3n+22)^{1/2}(\ln (L/l_0))^{1/2}L^{-3}\left[1-\frac{c}{2}\frac{\ln|\ln (L/l_0)|}{|\ln (L/l_0)|}+\ldots\right].
\end{equation}
Determining the constant $c$ requires  to  extend  the analysis in 
 Ref.~\cite{eisen:88}  which is left for future research.
 Gaussian fluctuations 
give contributions of at least $O(v^0)$ which are therefore
of order  $L^{-3}$  
and thus subdominant (see  Eq.~(\ref{eq:as})).  
We compare Eq.~(\ref{eq:as}) for $n=2$ with 
the data obtained 
 by Garcia and Chan~\cite{garcia:02:0} for   their
experimental value  of  $L\approx 520 $\AA  ~and for  $l_0\approx 1.3$ \AA,
 the experimental value of the
correlation length amplitude $\zeta_0=\zeta(t)/|t|^{-\nu_t}$ with $\nu_t=1$ 
 for {\it concentration} fluctuations
below $T_t$ in the superfluid phase \cite{leiderer}. 
For these values Eq.~(\ref{eq:as}) predicts
\begin{equation}
\label{eq:2a_3}
\vartheta_t\equiv f_C^t L^{3} \approx 6.96
\end{equation}
whereas 
 $\vartheta_t^{exp}=8.4\pm 1.7$.
 The value of the theoretical function $\vartheta_t$ at $T_t$, with $l_0$ between 
1 and 2 \AA , is in  reasonable agreement with the measured $\vartheta^{exp}_t$.
In order to  extract the actual
value of the {\it universal} Casimir amplitude (i.e., the numerical prefactor
$0.54 \sqrt{ 28}=2.86$  in Eq.~(\ref{eq:as})) the experimental data call for
 a re-analysis
based on the functional form given by Eq.~(\ref{eq:as}), which renders
 the comparison  independent of the choice for $l_0$, and requires to take
 into account the correction terms given in Eq.~(\ref{eq:as}).
We want to emphasize that the 
 tricritical Casimir force offers the opportunity
to observe the so far experimentally elusive logarithmic corrections 
associated with tricritical phenomena.
We note, that  {\it at tricriticality} the Casimir force $f_C^t(L\to\infty)$
 dominates over the background dispersion forces. This differs from the
case of {\it critical} Casimir forces  for which both contributions 
decay with
 the same power law.
It is interesting  that the Casimir amplitude for the present $(+,O)$  BC
is the same as for $(+,+)$ BC considered in Ref.~\cite{ritschel}.

\subsection{Logarithmic corrections to  the scaling function}
\label{subsec:23}

The scaling properties  of the Casimir force  follow  from the 
renormalized  finite-size contribution to the 
excess free energy (Eq.~(\ref{eq:1a1})). For carrying out   the
 renormalization procedure of this quantity two aspects are relevant.
First, for the  film geometry, 
the width $L$ of the system is not  renormalized
\cite{privman}.
Second,  in the renormalized (R) finite-size contribution to the free energy
 $\delta f(L)$ (see the text before Eq.~(\ref{eq:1b})) the contributions from the
additive counter terms cancel  and one has ~\cite{diehl:86:0,footnote}:
\begin{equation}
\label{eq:23_1}
\delta f^R(r,u,v;\mu,L)=\delta f(r_0,u_0,v_0;L)
\end{equation}
where  the bare quantities $u_0, r_0$, and $v_0$ are expressed in terms 
of renormalized ones  $r, u$, and $ v$; 
$\mu$ is an arbitrary momentum scale.
Since we are not considering  correlation functions  at the surface, all 
renormalization factors $Z$ are the same as those in 
 the  bulk \cite{diehl:86:0,eisen:88}:
\begin{equation}
\label{eq:23_2}
r_0=Z_{r}r+u^2\mu^{-2\epsilon}P, \quad u_0=Z_uu, \quad v_0=2\pi^2Z_vv,
\end{equation}
where the dimensions of the coupling constant are
$[r_0]=\mu^2, [u_0]=\mu^{1+\epsilon}$ and $[v_0]=\mu^{2\epsilon}$.
 Explicit perturbative results for the tricritical bulk
renormalization functions $Z_r, P , Z_u$, and $Z_v$ are 
known (see, e.g.,  Refs.~\cite{LawSar,eisen:88}).
From Eq.~(\ref{eq:23_1}) the RG equation can be derived 
in a standard fashion by  exploiting the fact that  
 $\delta f(r_0,u_0,v_0;L)$ is independent of $\mu$. 
Because in  Eq.~(\ref{eq:23_1}) 
 there are no additive renormalization
terms  it follows that $\delta f^R(L)$ 
satisfies the following  homogeneous RG equation \cite{diehl:86:0}:
\begin{equation}
\label{eq:23_3}
\left[\mu\partial\mu +\sum_{\kappa=r,u,v}\beta_{\kappa}\partial_{\kappa}\right]\delta f^R(L)=0
\end{equation}
where $\beta_{\kappa}(r,u,v;\epsilon)\equiv \mu\partial_{\mu}|_0\kappa$ and
 $\partial_{\mu}|_{0}$ denotes derivatives  with respect to $\mu$
at fixed bare interaction 
constants for   $\kappa=r,u,v$. 
The RG equation is solved by using the method of characteristics
 (see, e.g., Ref.~\cite{amit}):
\begin{equation}
\label{eq:23_4}
\delta f^R(r',u,v,\mu;L)=\delta f^R({\bar r'}(\ell),{\bar u}(\ell),\bar{v}(\ell);\mu \ell;L)
\end{equation}
where $\ell$ is again a dimensionless spatial rescaling factor,
 ${\bar \kappa}(\ell)$
 are the running coupling constants with the
 initial condition
${\bar \kappa}(1)=\kappa$,
and due to the form of the renormalization 
 of $r_0$ (see Eq.~(\ref{eq:23_2})) the new variable $r'$ is 
given by  \cite{LawSar,eisen:88} 
\begin{equation}
\label{eq:nower}
 r'=r+w(v,\mu)u^2.
\end{equation}
For an explicit expression of $w(v,\mu)$ see Refs.~\cite{LawSar,eisen:88}.
Equation~(\ref{eq:23_4})  summarizes the RG transformation 
and the non-renormalization of $L$.
 Using  dimensional analysis one obtains
\begin{equation}
\label{eq:23_5}
\delta f^R(r',u,v,\mu;L)=(\mu \ell)^{(d-1)}\delta f^R\left(\frac{{\bar r'}(\ell)}{(\mu \ell)^2},\frac{{\bar u}(\ell)}{(\mu \ell)^{4-d}},\frac{\bar{v}(\ell)}{(\mu \ell)^{2(3-d)}};1,L\mu\ell\right).
\end{equation}

The desired asymptotic scaling  behavior of $\delta f^R$ 
follows by substituting on the rhs of Eq.~(\ref{eq:23_5})
the appropriate fixed-point values for the 
running coupling constants ${\bar r}', {\bar u},$  and ${\bar v}$.
 The infrared stable fixed point lies at  
 $v^{\ast}=(240/(3n+22))\epsilon +O(\epsilon^2)$  \cite{eisen:88}.
Upon approaching  the upper
critical dimension   $v^{\ast}\to 0$ and  for $\epsilon\to 0$
 the relevant  logarithmic
corrections to the classical exponents are generated by the
flow of the coupling constants under the RG transformation $\ell\to 0$.
In the limit $\ell\to 0$, ${\bar v}(\ell)$ is given by Eq.~(\ref{eq:2a_2}). The
 running variables ${\bar r'}(l)$ and ${\bar u}(l)$ can be written as 
 ${\bar r'}(\ell)=E_{r}(\ell;v)r'$ and ${\bar u}(\ell)=E_u(\ell;v) u$. 
A straightforward 
analysis \cite{LawSar,eisen:88} shows that
$E_r(\ell;v)\to const $ and $E_u(\ell;v)\sim |\ln \ell|^{-2(n+4)/(3n+22)}$
for $\ell\to 0$. Choosing $\mu=1/l_0$,  $\mu\ell L=\ell(L/l_0)=1$, and omitting
the  constant factor
$E_r$ we obtain the following  scaling form for $\delta f$:
\begin{equation}
\label{eq:23_6}
\delta f^R(r',u,v,\mu;L)=L^{-2}\delta f^R(r'L^2,uL|\ln (L/l_0)|^{-2(n+4)/(3n+22)},|\ln (L/l_0)|^{-1};1,1).
\end{equation}
Due to  Eq.~(\ref{eq:1a}) the scaling form for the  Casimir force
follows from Eq.~(\ref{eq:23_6}) as: 
\begin{equation}
\label{eq:23_7}
f_C(r',u,v;L)\simeq L^{-3}\theta (r'L^2,uL|\ln (L/l_0)|^{-2(n+4)/(3n+22)},|\ln (L/l_0)|^{-1}).
\end{equation}
The scaling function $\theta$ is given in terms of $\delta f^R(z_1,z_2,z_3,1)$ as
$\theta=2\delta f^R+2z_1(\partial \delta f^R/\partial z_1)-
z_2(\partial \delta f^R/\partial z_2)$. 
The higher-order terms neglected in Eq.~(\ref{eq:23_7}) are of the  form
$L^{-3}(\ln (L/l_0))^{-1}(2(n+4)/(3n+22))z_2(\partial \delta f^R/\partial z_2)+L^{-3}(\ln (L/l_0))^{-1}z_3(\partial \delta f^R/\partial z_3)+L^{-3}(-1+2c(\ln|\ln (L/l_0)|)/(\ln^3|(L/l_0)|)(\partial \delta f^R/\partial z_3)$.
The third term in the latter expression stems
 from the correction to $z_3$ (see Eqs.~(\ref{eq:23_5}) and (\ref{eq:2a_2})).

At the upper critical dimension the asymptotic critical behavior
obtained from the perturbative  RG calculations 
within the  Gaussian approximation is expected to  be exact. However,
at the lowest order, often referred to as renormalized mean-field
theory (RMF) -- which yields the free energy correctly with the leading logarithms --
one neglects the contributions stemming from the Gaussian fluctuations
and replaces the scaling function by its mean-field-like form but 
with the rescaled arguments.

Applying this reasoning  to the free energy   we use
the mean-field result given by Eqs.~(\ref{scaltriC}) and (\ref{implicit})
with $r_0$ replaced in favor of $r'$ according to  Eq.~(\ref{eq:nower})
with  $w({\bar v}(\ell),\mu (\ell))\to const$ as $\ell\to 0$, 
$u_0$ replaced by  $u|\ln (L/l_0)|^{-2(n+4)/(3n+22)}$,
and $v_0$ replaced by $((240 \pi^2)/(3n+22))|\ln (L/l_0)|^{-1}$
to obtain at  lowest order:
\begin{equation}
\label{scaltriC1}
f^{RMF}_C\simeq \left(\frac{3n+22}{8\pi^2/3}\right)^{1/2} \left(\ln (L/l_0)\right)^{1/2}L^{-3}\vartheta^{MF} \left(r'L^2,uL|\ln (L/l_0)|^{-2(n+4)/(3n+22)},|\ln (L/l_0)|^{-1}\right).
\end{equation}
 
In the following we want
 to compare the behavior of the  MF and RMF expression for  the Casimir force.
As we have already stressed before,  $f_C$ calculated within the MF approach
depends on the non-universal and dimensionless parameter $v_0$ (see Eq.~(\ref{scaltriC})). 
Upon  comparing with  experimental data this parameter can be used 
to fit the amplitude of the Casimir force, because $v_0^{-1/2}$
appears (albeit not exclusively) as a prefactor of the scaling function. 
The factor  $v_0^{-1/2}$, which 
 multiplies
 the coupling constant $u_0$ (see the text after Eq.~(\ref{implicit})), 
is  absorbed  in the definition of the  scaling variable.

In Fig.~\ref{fig:0}  we have plotted  two curves:
(1)
 ${\bar \vartheta}^{MF} (r_0L^2=0,y^{MF})=f_CL^3(v_0/90)^{1/2}$
as  a function of
 $y^{MF}=(5/(2v_0))^{1/2}u_0L$ (see Eq.~(\ref{implicit})). 
 Here, the non-universal factor  $v_0^{-1/2}$ is
 absorbed  in  the definitions of the scaling function and of
 the scaling variable. As  already mentioned  before  $u_0\sim t$ if $r_0=0$, 
so that  $u_0L \sim tL$.
(2)  
$f_CL^3\equiv {\bar \vartheta}^{RMF}(0,y^{RMF})= (28/(8\pi^2/3))^{1/2}(\ln (L/l_0))^{1/2}\vartheta^{MF} (0,y^{RMF})$  (for $n=2$), where  
 $y^{RMF}= uL(\ln (L/l_0))^{1/14}$.
 Here, renormalization fixes the
amplitude of the Casimir force replacing  the non-universal prefactor
$\left(\frac{v_0}{90}\right)^{-1/2}$ (see Eq.~(\ref{scaltriC}))
  of
the scaling function by the amplitude and the logarithmic correction to the
$L$ dependence.  
The scaling variable $y^{RMF}$ includes the logarithmic correction
$|\ln (L/l_0)|^{-2(n+4)/(3n+22)}$  to $u$ and
the additional logarithmic term $|\ln (L/l_0)|^{1/2}$ stemming  from
the factor  $(5/(2v_0))^{1/2}$
(see $y^{MF}$ and Eq.~(\ref{implicit})). The numerical factor  $7/(24\pi^2)$
 has  been included into the  definition of  $u$. 
 For comparison with  experimental data  this factor
can be combined  with the non-universal constant of proportionality
between $u$ and $t$.
For the plot we have
chosen the experimental value for $L/l_0$, i.e., 520 \AA /1.3 \AA. 
The shapes of both scaling functions  are
 similar but the RMF
result gives the correct value for  
the Casimir amplitude and the  correct $L$-dependence of the scaling function. 
This should be helpful for interpreting 
 experimental  data obtained  for different film  thicknesses.

In Fig.~\ref{fig:0a} we  show  the corresponding results for $u=0$ so that
$r'=r$ (see Eq.~(\ref{eq:nower})), $r\sim t$, and $rL^2 \sim gL^2$.
We find that for  $u=0$ both scaling functions decay much faster
to zero than for  $r=0$.

For $r_0=0$ one has $u\sim t$ so that, up to the logarithmic corrections, 
the scaling function 
 ${\bar \vartheta}^{MF} (0,y^{MF})$  should correspond
to the  experimental curve  $\vartheta(tL)$  
in  Fig.~\ref{fig:3a} for the tricritical concentration \cite{garcia:02:0}. 
(We note  that the
 argument of the experimental curve is given  in units of \AA .)
 The solid line in this figure represents 
${\bar \vartheta}^{MF} (0,y^{MF})$ suitably adjusted with respect to
  the parameter $v_0$  such that the Casimir amplitude and the 
position of the maximum  equal the experimental ones \cite{garcia:02:0}.

\section{Vectoralized Blume-Emery-Griffiths Model.}
\label{sec:3}

Based on the motivation provided in the Introduction, in this section we extend
the VBEG model to the film geometry and study  $^3$He-$^4$He mixtures.

\subsection{The model}
\label{subsec:1}
We consider a three-dimensional slab of a simple  
cubic lattice consisting of ${\bar L}$ parallel  $(100)$ lattice 
layers with lattice  spacing $a$ so that $L={\bar L}a$.
 Each layer has ${\bar A}=A/a^2$ sites,
labeled $i, j, \ldots$, which are   associated with an occupation  
variable  $t_i=0,1$
and a phase $\theta _i$ $( 0\le \theta _i< 2\pi)$ which mimics
 the phase of the $^4$He  wave function and thus renders
 the XY bulk universality class ($n=2$).
A $^3$He ($^4$He) atom at site $i$ corresponds  to $t_i=0 (1)$
so that in the bulk $X=1-\langle t_i\rangle$ is the $^3$He concentration.
Unoccupied sites are not allowed so that
 the model does not exhibit a vapor phase. Accordingly this model does 
not allow for the occurrence of a tricritical end point.
 However, we expect that
the universal properties we are interested in are the same for 
tricritical points and tricritical end points.
 The Hamiltonian consists of bulk and 
surface contributions ${\cal H}={\cal H}_b+{\cal H}_s$  with 
\begin{equation}
\label{eq:ham}
{ \cal H}_b=-J\sum _{\langle ij\rangle}t_it_j\cos (\theta _i-\theta
_j) -K\sum _{\langle ij\rangle}t_it_j+\Delta\sum _it_i,
\end{equation}
where    the first two sums run over  nearest-neighbor pairs and the
last one  is over all lattice sites, except those at the surface.
In this  lattice gas model of   $^3$He-$^4$He binary mixtures  the 
coupling constant $ K$ and the field $\Delta$ are related to the effective
$^{\alpha}$He-$^{\beta}$He interactions 
  $K_{\alpha \beta}$  (see, e.g., Ref.~\cite{bell:89:0}),
\begin{equation}
\label{eq:parK}
K=K_{33}+K_{44}-2K_{34},
\end{equation}
and to the chemical potentials $\mu _3$ and $\mu _4$  of  $^3$He and $^4$He,
 respectively,
\begin{equation}
\label{eq:pardelta}
\Delta=\mu _3-\mu _4+2q(K_{33}-K_{34}),
\end{equation}
where  $q$ is  the coordination number
 of the lattice ($q=2d$, where $d$ is  the 
spatial dimension of the system; $q=6$ in the present case).
 In the liquid the effective interactions 
 $K_{\alpha \beta}$ are different for different $\alpha$
 and $\beta$ due to the differences in  mass and of statistics  between $^3$He
 and $^4$He atoms.

The properties of the model described by the bulk Hamiltonian ${\cal H}_b$
 have been  studied within
 MFT and by Monte Carlo simulations 
in $d=3$  \cite{maciolek:04:0}.
In contrast to its
two-dimensional version, for which  there  is
 no true tricritical point for any value of the model parameters, in $d=3$
for reasonable values of the interaction parameters
the resulting phase diagram resembles that observed
experimentally for $^3$He-$^4$He mixtures, for which phase
separation occurs as a consequence of the superfluid transition
(see Fig.~\ref{fig:1}). 
The form  of the surface Hamiltonian ${\cal H}_s$ should capture the phenomenon
of superfluid film formation near a wall in 
 $^3$He-$^4$He mixtures \cite{laheurte:78:0} which generates
 an effective repulsion of $^3$He atoms by the wall. 
The van der Waals interactions between the wall  and $^3$He or
$^4$He atoms are equal.  However, $^3$He atoms  occupy  a larger volume 
because of their  larger zero-point motions. This gives rise 
to the preferential adsorption of $^4$He atoms
at the substrate-fluid interface, which may induce a local superfluid ordering
and  an enrichment of $^3$He near the 
opposing fluid-vapor interface.
Here  we choose
 the following  form for ${\cal H}_s$:
\begin{equation}
\label{eq:surfham}
{ \cal H}_s=\delta \Delta^{(l)}\sum _i^{(l)}t_i+\delta \Delta^{(r)}\sum _i^{(r)}t_i,
\end{equation}
where the first  sum runs over the sites of the first 
 layer  and the second over those in  the $L$-th layer of the lattice.
The differences
 $\delta \Delta^{(l)}\equiv \Delta^{(l)}-\Delta$ 
and $\delta \Delta^{(r)}\equiv\Delta^{(r)}-\Delta$  are  measures of  
 the relative preferences
of $^4$He atoms for the two  surfaces 
 such that $\delta \Delta^{(l)}<0$ corresponds
to the preference of $^4$He atoms for the 
solid substrate. 

\subsection{Mean-Field Theory}
\label{subsec:2}

\newcommand{\Tr}{{\rm Tr}}

We have studied the above model for the film geometry 
within  mean-field theory.
We have employed the variational method  based on approximating the 
total equilibrium  density distribution  by a product  of local site
 densities
$\rho _i$  (see, e.g., Ref.~\cite{book}).
The corresponding variation theorem for the free energy reads
\begin{equation}
\label{eq:var}
F\le F_{\rho}=\Tr(\rho{\cal H})+(1/\beta)\Tr(\rho \ln\rho),
\end{equation}
where $F$ is the exact free energy and $F_{\rho}$ is an approximate free energy
 associated with the density distribution  $\rho$; $\beta =1/(k_BT)$.
The minimum of $F_{\rho}$  with respect to  $\rho$ subject to 
the constraint $\Tr\rho =1$ is attained  for the equilibrium density
 distribution
$\rho=e^{-\beta {\cal H}}/\Tr(e^{-\beta {\cal H}})$.
Within mean-field theory the density distribution  in the film geometry
is approximated by
\begin{equation}
\label{eq:denmat}
\rho=\rho_0={\bar A}^L\prod_{i=1}^L\rho_i,
\end{equation}
i.e., the density distribution  is constant within each  layer parallel to
the surfaces but  
varies from layer to layer.
We treat the local layer  density $\rho _i$ as a variational  ansatz,
and the  best functional form in terms of $t_i$ and $\theta _i$ is 
obtained by minimizing  $ F_{\rho _0}/{\bar A}+\eta \Tr(\rho _i)$ with respect
 to $\rho _i$ and with  $\eta $ as a Lagrange multiplier in order to implement
$\Tr \rho=1$. This leads to
\begin{equation}
\label{eq:solden}
\rho _i=e^{-\beta h_i}/\Tr(e^{-\beta h_i})
\end{equation}
where  $h_i$ is the single-layer  mean field  given by
\begin{eqnarray}
\label{eq:h}
 h_i=&-&J(M^{(1)}_{i-1}+q_{||}M^{(1)}_{i}+M^{(1)}_{i+1})t_i\cos\theta _i-J(M^{(2)}_{i-1}+q_{||}M^{(2)}_{i}+M^{(2)}_{i+1})t_i\sin \theta _i  \nonumber \\
&-&K(Q_{i-1}+q_{||}Q_i+Q_{i+1})t_i+\Delta^{(i)}t_i,
\end{eqnarray}
where $\Delta^{(i)}=\Delta$ for $i\ne 1, {\bar L}$, and 
$\Delta^{(i)}=\Delta^{(l)}  (\Delta^{(r)})$ for $i=1 ({\bar L})$.
We have introduced  the following order parameters:
\begin{equation}
\label{eq:q}
Q_i\equiv 1-X(i)=\Tr(t_i\rho _i)
\end{equation}
and 
\begin{equation}
\label{eq:m}
M^{(1)}_i= \Tr(\rho_it_i\cos \theta _i), \qquad M^{(2)}_i= \Tr(\rho_it_i\sin \theta _i).
\end{equation}
$Q_i=\langle t_i\rangle$ corresponds to the concentration profile of  $^4$He,
 $X(i)=1-\langle t_i\rangle$ to the concentration profile of 
$^3$He, and $M^{(1)}_i, M^{(2)}_i$
are the components of the two-component  superfluid OP profile
 ${\bf M}_i$.  $q_{||}$ is the in-layer coordination 
number while each site  (but not in the first and last layer) is connected 
to $q'$ atoms in each  adjacent layer and  $q=q_{||}+2q'$ is 
 the coordination number in the bulk  of the lattice.  Within our model
$q'=1$ and $q_{||}=2(d-1)$.
This  yields  the following  set of self-consistent 
 equations for the OP ${\bf M}_i=(M^{(1)}_i,0)\equiv (m_i,0)$
in the $i$th layer :
\begin{equation}
\label{eq:set1}
Q_i=I_0(\beta Jb_i)/\left(e^{-\beta(Ka_i-\Delta^{(i)})}+I_0(\beta Jb_i)\right),
\end{equation}
and
\begin{equation}
\label{eq:set2}
m_i=I_1(\beta Jb_i)/\left(e^{-\beta(Ka_i-\Delta^{(i)})}+I_0(\beta Jb_i)\right).
\end{equation}
$I_0(z)$ and  $I_1(z)$ are the modified Bessel functions of the first kind, $T$ is the temperature.
We have introduced
\begin{equation}
\label{eq:bl}
b_i\equiv m_{i-1}+q_{||}m_i+m_{i+1} \qquad \mbox{for}\qquad  i\ne 1, {\bar L},
\end{equation}
 $b_1\equiv q_{||}m_1+m_2$, and  $b_{\bar L}\equiv m_{{\bar L}-1}+q_{||}m_{\bar L}$,
and analogously
\begin{equation}
\label{eq:al}
a_i\equiv Q_{i-1}+q_{||}Q_i+Q_{i+1} \qquad \mbox{for} \qquad i \ne 1, {\bar L},
\end{equation}
  $a_1=q_{||}Q_1+Q_2$, and $a_{\bar L}=Q_{{\bar L}-1}+q_{||}Q_{\bar L}$.
The coupled sets of equations for $Q_i$ and $m_i$ are solved numerically by standard methods of multidimensional root finding. 
The equilibrium  solution minimizes the free energy per number of lateral lattice sites
 ${\cal F}\equiv F_{\rho _0}/{\bar A}$:
\begin{eqnarray}
\label{eq:freeen}
f &=&\sum_{i=2}^{{\bar L}-1}\left[\frac{J}{2}(m_{i-1}m_i+q_{||}m_i^2+m_{i+1}m_i)+\frac{K}{2}(Q_{i-1}Q_i+q_{||}Q_i^2+Q_{i+1}Q_i)\right] \nonumber \\
&+&k_BT\sum_{i=1}^{{\bar L}}\ln (1-Q_i)+f_1+f_2,
\end{eqnarray}
where
\begin{equation}
\label{eq:phi1}
f_1=\frac{J}{2}(q_{||}m_1^2+m_{2}m_1)+\frac{K}{2}(q_{||}Q_1^2+Q_{2}Q_1)
\end{equation}
and 
\begin{equation}
\label{eq:phi2}
f_2=\frac{J}{2}(m_{{\bar L}-1}m_{\bar L}+q_{||}m_{\bar L}^2)+\frac{K}{2}(Q_{{\bar L}-1}Q_{\bar L}+q_{||}Q_{\bar L}^2).
\end{equation}
The above equations  neglect the  helicity, i.e.,  
${\bf M}_i=(M^{(1)}_i,0)\equiv (m_i,0)$. In general
the helicity  might be 
 non-zero because the  BC for the superfluid OP are
 effectively non-symmetric, i.e., ${\bf M}_1\ne 0$ whereas  ${\bf M}_L=0$ 
so that  the superfluid OP can in principle  rotate 
across the film. The relevance of the helicity on the Casimir force
 will be analyzed elsewhere.

\subsection{Results for $^3$He-$^4$He mixtures}
\label{subsec:3}

First,  we have analyzed
 the semi-infinite system. Close to the line of {\it bulk} critical points
 we have found  a
 higher $^4$He concentration near the  
  surface (chosen to be  the left side of the system),
 which  induces   a local superfluid ordering.
 By varying $T$ and
$\Delta$ one obtains a   line of continuous
 {\it surface} transitions  corresponding to the onset of the formation
of  this  superfluid film near the wall; it
meets the line  of bulk critical points
at a so-called  special transition point, the  position of which
depends on the value
of    $\Delta ^{(l)}$ (see Fig.~\ref{fig:1}). These  findings are in  
 agreement with the results
 of a Migdal-Kadanoff analysis~\cite{peliti:85:0}.

In the film geometry the Casimir force $f_C$ (Eq.~(\ref{eq:1a}))
 is obtained by   calculating $f^{ex}({\bar L})$ (see Eq.~(\ref{eq:1a1}))
for  ${\bar L}$ and ${\bar L}+1$  and taking the difference.
(Note that in the lattice model $f$ is the total free energy of the film
per number ${\bar L}{\bar A}$ of lattice sites and $f_b$ is the bulk free
energy density per ${\bar L}{\bar A}$.
Accordingly $f^{ex}({\bar L})=(f-f_b){\bar L}/(k_BT_t)$, $f_C=-\partial f^{ex}/\partial {\bar L}$, as well as $\vartheta=f_C{\bar L}^d$ with $d=3$ 
near tricriticality and $d=4$ near the $\lambda$-transition are dimensionless.
In order to avoid a clumsy notation we do not introduce different symbols for
the lattice and the continuum versions of the free energies.)
Figure~\ref{fig:3} summarizes our result for a  film of 
thickness ${\bar L}=20$,
$K/J=0.5$, $\Delta^{(l)}/ J=-3$, and 
$\Delta ^{(r)}=\Delta _t/J\simeq 0.61$, which is the tricritical bulk  value.
 Such a choice of the surface 
coupling constants corresponds to  non-symmetric BC and is consistent with
the assumption made in Ref.~\cite{garcia:02:0} for the concentration profile
across the wetting film, whereupon at the interface with the vapor 
 the $^3$He concentration  takes the bulk value.
 For temperatures above
the bulk  coexistence line  at first-order demixing transitions 
$f_C$ is calculated along the
 thermodynamic  paths indicated in Fig.~\ref{fig:1} which 
 correspond to  fixed 
$^3$He concentrations $X$. Our  selection of $X$
covers the tricritical region as well as the crossover to the
 critical superfluid behavior of pure $^4$He, i.e., $X=0$. 
In order to calculate the force at a
fixed value  $X_0$  we first determine  $\Delta (X=X_0,T)$  by solving 
the two coupled self-consistent equations for
 $Q(\Delta,T)=1-X$ and $M(\Delta,T)$ in the  bulk 
(Eqs.~(12) and (13) in Ref.~\cite{maciolek:04:0}). 
For each temperature along the  thermodynamic paths indicated
in Fig.~\ref{fig:1} we solve Eqs.~(\ref{eq:set1}) and 
(\ref{eq:set2}) with this value  $\Delta (X=X_0,T)$.
This renders the  profiles $Q(l)$ and $m(l)$ and allows us to
 calculate the free energy from 
Eq.~(\ref{eq:freeen}). When upon lowering the temperature 
the paths of constant $X$ reach  the
coexistence line of two-phase coexistence (see Fig.~\ref{fig:1})
 we continue our calculations along the coexistence 
line, infinitesimally on the superfluid branch of  bulk coexistence. In
Fig.~\ref{fig:3} this leads to the  full line  for  $T<T_t$, i.e., $y<0$.

Contrary to the  LG  model,  for the present
 microscopic model it is natural to express the properties of the system
as  functions  of the experimental
thermodynamic  fields $t$ and $(\Delta -\Delta_t)/(k_BT_t)$ or the scaling fields
$t$ and $g$ (see Eq.~(\ref{scalfields})).
Accordingly, we present our results for the Casimir force  in terms 
of the  scaling function defined through the relation
 $\vartheta \equiv {\bar L}^3 f_C$ as  a function of only a single 
  scaling variable 
 ${\bar y}\equiv t{\bar L}^{1/\nu}=((L/a)/(\xi/\xi_0^+))^{1/\nu}$.
$\xi^+_0={\bar \xi}^+_0a$ is the  amplitude of the order parameter
 correlation length $\xi=\xi^+_0t^{-\nu}={\bar \xi}a$ 
above $T_t$ and $\nu(d=3) =1$.
The second relevant scaling variable  $x\equiv g{\bar L}^2$
  also  varies along a  path
  of fixed $^3$He concentration (see Fig.~\ref{fig:1a}) and a proper
scaling description has to  account for it. However,
in order to be able to compare our results
with the presentation of the corresponding 
 experimental ones \cite{garcia:02:0},
we follow Ref.~\cite{garcia:02:0} where the variation of $x$
 has been neglected.  As can be inferred from the phase diagram
 in
Fig.~\ref{fig:1a},  the $g$-components of the paths  $X=const$ 
 in the phase diagram are smaller than the $t$-components, so that
the form of the scaling function for these paths are expected  be close 
to   $\vartheta (x=0,y)$. Also  experimentally the variation of
 the scaling variable $g$ along  the path of 
fixed $X$  cannot be  determined easily.

Near the tricritical point  paths of constant $X$ cross
three different phase transition lines: the surface transition line,
the line of bulk critical points,  
and the line of first-order phase coexistence.
As shown in Fig.~\ref{fig:3}, close to the  surface transition $f_C$ is 
small  and
this transition does not leave a visible trace 
in its  behavior.  $f_{C}$ remains small up  to the coexistence line 
or to the line of bulk critical points
 for $X>X_t$ or  $X<X_t$, respectively. There  it increases very steeply
and for $^3$He concentrations $X < X_t$ 
upon crossing   the line of bulk critical points there is  a break in slope 
(see the dots in Fig.~\ref{fig:3}) giving  rise to the formation
of  shoulders which are similar to those observed 
experimentally \cite{garcia:02:0}. When   $T$  reaches the
temperature of first-order phase separation,  $f_{C}$ is given by 
  the curve (full line for $y<0$ in Fig.~\ref{fig:3}) 
common to all  values of $X$. These 
curves of constant $X$  meet the full line  with different slopes.

The aforementioned common curve exhibits a pronounced maximum 
below $T_t$ at $y\simeq -0.74$ and gradually  decreases to zero for
 $y\to -\infty$. The properties of the Casimir force in this temperature
region  can be attributed to purely interfacial effects.
Indeed, we observe that below $T_t$  both the concentration 
 and the superfluid OP  
profile  corresponding to this common curve display
 an interface-like structure  separating two 
domains of the coexisting bulk phases
(see the case $t=-0.0633$ in  Fig.~\ref{fig:2}).  
This film  phase is soft with respect to shifts of the interface
position and is similar to the one occurring in  Ising-like
 films with opposite  BC~\cite{parry:92:0} for temperatures below the
bulk critical temperature but above the wetting temperature of the
confining walls, in which case  the Casimir force is repulsive
with a pronounced maximum occurring  below the bulk critical
temperature \cite{stecki}. In general  a
positive sign of  the force can be regarded
as a consequence of entropic repulsion~\cite{fisher:86:0}.
Typically the maximum of the force occurs at that temperature $T$ at 
 which  the
interfacial width, which is proportional to the bulk correlation
length $\xi$ of the order parameter, becomes comparable with the width
$L$ of the film.
In the present case both the  concentration
and the superfluid OP  profile  contribute to the free energy
and hence to the Casimir force.
Their interfacial widths are proportional to correlation
length $\zeta$ associated with concentration fluctuations and to the
OP correlation length $\xi$, respectively.
As can be seen  from Figs.~\ref{fig:2} and \ref{fig:profopp}, within
MFT these interfacial widths and therefore $\zeta$ and $\xi$ are comparable.
Accordingly, by analogy with Ising-like systems~\cite{parry:92:0}
we expect that within MFT the maximum of the force  occurs when $\xi$
(or, equivalently, $\zeta\simeq \xi$) is of the order of
$L$, which is actually consistent with what is observed in
Fig.~\ref{fig:3}, where the maximum of the scaling function
is located at $y\simeq -1$.
We may expect that also in the actual  system the occurrence of the maximum
of the Casimir force below the tricritical point
can be attributed to such  interfacial effects. However, 
since the correlation length of the superfluid
OP $\xi=\infty$ in the superfluid phase
 it is not yet clear which length scale governs the interfacial width 
of the superfluid OP profile in the 'soft mode' phase  below $T_t$
and hence  what length scale determines  the position  of the force  maximum.

For $X \lesssim X_t-0.05$ we observe a  crossover to the critical superfluid
behavior of pure $^4$He and a  gradual formation of a second, less pronounced 
local maximum located slightly below the line of bulk critical points
 ($y>0$ in Fig.~\ref{fig:3}).
 This local maximum decreases 
upon departing  from $X_t$ and finally $f_C$ becomes vanishingly small
 along paths which
cross the line of bulk critical points
above the special transition $S$ (see Fig.~\ref{fig:1}). This is expected,
because  above $S$ there is no longer a superfluid film formation near the
solid substrate for thermodynamic states corresponding to 
the bulk ``normal'' phase of a fluid close to  the line of bulk
 critical points.
This means that the superfluid OP in the film is identically
 zero up to  the line of bulk critical points
and   the BC effectively turn into the type $(O,O)$ for which 
$f_{C}$ vanishes within MFT. (For (O,O) BC fluctuations beyond MFT generate
an attractive Casimir force $f_C<0$ \cite{krech:91}.)
For lower $T$, $f_{C}$ increases steeply upon  approaching bulk coexistence
revealing that   interfacial effects associated with the 
'soft mode'  lead  to a much stronger Casimir effect than the 
critical fluctuations  near the  line of bulk critical points.

\section{Results for pure $^4$He}
\label{sec:4}
\subsection{The limiting case of the VBEG model}
\label{subsec:4-1}

In this section we  consider the limiting case  $\Delta\to -\infty$
in which  all lattice sites are
occupied, i.e., $t_i\to 1$. In this case 
 the first term of the bulk Hamiltonian ${\cal H}_b$ in Eq.~(\ref{eq:ham})
corresponds to  the classical XY model 
(the planar rotator model) for pure $^4$He and therefore, as far as
 the bulk contribution is concerned, 
the partition function 
of the VBEG model  reduces to that of the XY  model  up 
to a factor $e^{KzN}$ where $N$ is the
number of lattice sites. 
The corresponding  MFT equations for the  bulk OP
can be inferred from Eqs.~(\ref{eq:set1}) and (\ref{eq:set2}) 
with $m_i\equiv M$ yielding
\begin{equation}
\label{eq:lim}
Q=1, \qquad M=\frac{I_1(\beta qJM)}{I_0(\beta qJM)}
\end{equation}
for temperatures below  the bulk superfluid transition, which is
located   at $T_s(X=0)=T_{\lambda}=qJ/2$,
and $Q=1$, $M=0$ above $T_{\lambda}=T_s(X=0)$.
The scaling behavior of the free energy and of the Casimir force  close 
to this critical point (see below) is consistent with an upper 
critical spatial dimension 
$d^{\ast}=4$. The crossover to the tricritical behavior with $d^{\ast}=3$
and with tricritical exponents occurs only upon approaching
the tricritical point $A=(T_t/T_s(0)=2/3, X_t=1/3)$ (see Fig.~\ref{fig:1}).

In the slab geometry we take also the limits
$\Delta^{(l)}, \Delta^{(r)}\to -\infty$ which,  together 
with the absence  of external fields
coupling to the superfluid OP, lead  to
$(O,O)$ BC for the superfluid OP. 
Thus this  limiting case  allows us  to study  the Casimir force
for  wetting films  of pure $^4$He near the superfluid
 transition at $T_c=T_{\lambda}$. We remark that 
in the slab geometry the superfluid
transition is actually of the  Kosterlitz-Thouless 
type \cite{kosterlitz:80}.
However, this change of the character
of the transition  is not captured by MFT.
The corresponding set of  equations for the superfluid OP in  the $l$-th 
layer of the slab is:
\begin{equation}
\label{eq:lim1}
m_l=\frac{I_1(\beta Jb_l)}{I_0(\beta Jb_l)}, \qquad 
b_l\equiv m_{l-1}+q_{||}m_l+m_{l+1} \qquad \mbox{for}\qquad  l\ne 1, {\bar L},
\end{equation}  
where  $b_1\equiv q_{||}m_1+m_2$  and  $b_{\bar L}\equiv m_{{\bar L}-1}+q_{||}m_{\bar L}$.
The equilibrium free energy divided by the number $A$ of lattice sites within
one  layer  takes the form
\begin{eqnarray}
\label{eq:freeenlim}
f &=&\sum_{l=2}^{{\bar L}-1}\left[\frac{J}{2}(m_{l-1}m_l+q_{||}m_l^2+m_{l+1}m_l)\right]+\frac{J}{2}(q_{||}m_1^2+m_{2}m_1)+\frac{J}{2}(m_{{\bar L}-1}m_{\bar L}+q_{||}m_{\bar L}^2)\nonumber \\
&+&k_BT\sum_{l=1}^{{\bar L}}\ln (I_0(\beta Jb_l))
\end{eqnarray}
where $m_l, l=1,\ldots, {\bar L}$, are the solutions of Eq.~(\ref{eq:lim1}).
Solving   Eq.~(\ref{eq:lim1}) for different widths  of the film we have  found
that  the superfluid OP profile vanishes for temperatures
larger  than a certain   $T_c({\bar L})<T_s(X=0)=T_{\lambda}$ which
 can be identified
with the critical temperature  $T_c({\bar L})$ of the slabs.
Below $T_c(\bar{L})$ the corresponding Casimir force  turns out to be  negative (i.e., attractive)
 as  expected  for  $(O,O)$ BC pertinent to the case of pure $^4$He. 
The lattice calculations have been carried out for $d=3$ and are presented in
 terms of the scaling function  
 $\vartheta_0(y=\tau(L/\xi_0^+)^2) \equiv {\bar L}^d f_C$ with $d=4$
in accordance with MFT and $\tau\equiv(T-T_{\lambda})/T_{\lambda}$.
Within lattice MFT the actual space  dimensionality $d$ of the
lattice  does not influence 
the shape of the scaling function in the scaling limit $\xi(\tau>0) = \xi_0^+
\tau^{-\nu} \gg a$; indeed, it enters only into the non-universal amplitude
$\xi_0^+$ via the ratio $q'/q = (2 d)^{-1}$ between the bulk inter-layer
and the total site coordination numbers $q'$ and $q$, respectively.
$\vartheta_0$ has been calculated for  ${\bar L}=20, 40$, and $60$ 
and is  plotted in Fig.~\ref{fig:limsc} as a function of 
 $y\equiv \tau ( L/\xi_0^+)^{1/\nu}$ with the  MFT value   $\nu =1/2$. 
Exploiting the fact that within MFT $\xi(\tau<0)$ is finite we have
determined the amplitude 
$\xi_0^-$ of the correlation length $\xi(\tau<0) =
\xi_0^-(-\tau)^{-\nu}$ from the exponential approach 
of the  OP profiles towards  the corresponding bulk values
$m_b$ which are actually attained
in the middle of the film (see Fig.~\ref{fig:9}) 
at temperatures sufficiently below 
$T_s(X=0)$  (see, e.g., Fig.~21 in  Ref.~\cite{andrea}). 
The MFT universal amplitude ratio 
$\xi_0^+/\xi_0^- = \sqrt{2}$ then yields the estimate 
${\bar \xi}_0^+\simeq 0.41$  for the VBEG model on the lattice.
We emphasize here that scaling of the force data 
occurs only for surprisingly  thick  films, i.e., ${\bar L}\gtrsim
60$, as revealed clearly by the analysis presented in the next subsection.

\subsection{Comparison with the Landau-Ginzburg theory}
\label{subsec:4-2}

In Ref.~\cite{andrea} within MFT for the  $O(2)$ LG continuum  theory
(see Eq.~(\ref{eq:1}) with $v_0=0$) the order parameter profiles 
${\bf \Phi} = (m(z),0)$ in a slab with  $(O,O)$ BC have been calculated
analytically  (see Eqs.~(202) and (203) in Appendix D in
Ref.~\cite{andrea}). 
It turns out that as a function of the scaling
variable  $y=\tau(L/\xi_0^+)^{1/\nu} = r_0L^2$ 
(where $r_0\propto \tau$ is the coefficient appearing in Eq.~(\ref{eq:1}))
the mean-field OP profile $m(z)$ vanishes for $y\ge y_m \equiv -\pi^2$,
whereas it is nontrivial for $y<y_m$, breaking the original $O(2)$
symmetry. This occurs for temperatures below the
shifted critical point of the film 
which therefore corresponds to $y = y_m$ (see Ref.~\cite{fisher}).
In Fig.~\ref{fig:9} we compare
the  OP profiles (normalized by  the corresponding bulk values as to obtain
 universal scaling functions of $y$ and $z/L$) 
calculated within the  VBEG model (for  a lattice with $\bar L = 150$) and
within LG continuum theory  for a selection of values of the scaling
variable $y$. The agreement between the profiles is very good,
although the VBEG profiles exhibit a slight asymmetry with respect to
$z/L = 1/2$ which is due to the limited  numerical accuracy of 
the lattice  calculation.

The knowledge of the analytic expression for $m(z)$ allows one to
compute  the stress tensor (Eq.~(\ref{eq:1b}))
as a function of the scaling variable $y$:
\begin{equation}
\label{eq:MFF0}
{\cal T}_{zz} = \frac{1}{2} (m'(z=0))^2 =\left\{ 
\begin{array}{rl}
{\displaystyle \frac{A_m}{L^4} \frac{4 k^2}{(1+k^2)^2}\left(\frac{y}{y_m}\right)^2}, &
\mbox{for}\quad y < y_m = - \pi^2\,,\\
0, \phantom{000} & \mbox{for}\quad y \ge y_m\,,\\
\end{array}
\right.
\end{equation}
where $A_m = 3\pi^4/(2 u_0)$ and $k = k(y<y_m)$ 
is the real solution of the
implicit equation
\begin{equation}
\label{eq:MFF3}
 \frac{y}{y_m}=\frac{4}{\pi^2}(1+k^2)K^2(k)
\end{equation}
where $K(k)$ is the complete elliptic integral of the
first kind, such that $k(y=y_m)=0$ and $k(y\rightarrow-\infty) = 1$.
The stress tensor ${\cal T}_{zz,b}$ in the bulk, related to the {\it bulk} free energy
density $f_b(\tau)$, 
can be obtained from  Eq.~(\ref{eq:MFF0}) in the limit  
$L\rightarrow\infty$ at
fixed reduced temperature 
$\tau$, yielding ${\cal T}_{zz,b}(\tau < 0)
= A_m L^{-4} (y/y_m)^2$ (which is actually independent of $L$ due to
$y\propto \tau L^2$) and
${\cal T}_{zz,b}(\tau>0) = 0$. 
Accordingly, the Casimir force $f_C$ per unit area of
the cross section of the film and in units of $k_BT_\lambda$
is given by
$f_C = {\cal T}_{zz} - {\cal T}_{zz,b}$ and its scaling function
$\vartheta_0^{LG}=L^4f_C$ can be derived from the expressions
for  ${\cal T}_{zz}$ and ${\cal T}_{zz,b}$ discussed above:
\begin{equation}
\label{eq:MFF1}
\vartheta_0^{LG}(y) 
=\left\{ 
\begin{array}{rl}
- A_m {\displaystyle \left(\frac{1-k^2}{1+k^2}\right)^2 \left(\frac{y}{y_m}\right)^2 }
&\mbox{for}\quad y < y_m = - \pi^2\,,\\
- A_m {\displaystyle \left(\frac{y}{y_m}\right)^2} 
&\mbox{for}\quad y_m \le y < 0,\\
0\phantom{000} 
&\mbox{for}\quad y \ge 0.\\
\end{array}
\right.
\end{equation}
The independent calculation of $\vartheta_0^{LG}(y)$, recently
presented in Ref.~\cite{Zandi}, agrees with this expression.
%
At $y=y_m = -\pi^2$ the scaling function~(\ref{eq:MFF1}) exhibits 
a cusp singularity at which it attains
its minimum 
value $\vartheta_{\rm min}^{LG} \equiv \vartheta_0^{LG}(y_m)= - A_m<0$
where $A_m$ is given after Eq.~(\ref{eq:MFF0}). 
Within MFT the coupling constant $u_0$ and therefore $A_m$ remain undetermined. 
In order to compare the LG result with the VBEG results, accounting
also for corrections due to the finite size $\bar L$ of  the
latter, we introduce an adjusted scaling function
${\bar\vartheta}_0^{LG}(y)$ which is given by Eq.~(\ref{eq:MFF1})
with  $A_m = A_m(\bar L)$ and $y_m=y_m(\bar L)$  determined by a
best fit to the VBEG scaling function $\vartheta_0(y)$ calculated for 
lattices with $\bar L = 20$, $40$, and $60$. For all  values of $\bar L$
considered, ${\bar\vartheta}_0^{LG}(y)$ provides a very good fit to
the numerical data, as demonstrated  in Fig.~\ref{fig:limsc} for $\bar L =
60$. In the inset of Fig.~\ref{fig:limsc} we plot the functions
$A_m(\bar L)$ and $y_m(\bar L)$ obtained from the fit. According to
the results of the LG theory one expects $y_m(\bar L \rightarrow
\infty) = -\pi^2 \simeq -9.87$ (which is represented as a solid line
in the inset), and indeed the results of the VBEG model show the
correct trend, although finite-$\bar L$ corrections are still present
even for the largest lattice $\bar L=60$ considered here, with
$y_m(\bar L = 60) \simeq -9.31$. The amplitude
$A_m(\bar L)$ shows even stronger
corrections and indeed the value $A_m(\bar L
= 60) \simeq 2.45$ might underestimate the actual asymptotic value by
15-20\%.

Beyond MFT the renormalized coupling constant $u$ attains its fixed-point value
under RG flow which fixes the amplitude $A_m$ 
and the magnitude of the corrections to the scaling functions.
This would then allow a complete
numerical test with the scaling function $\vartheta_0$ of the VBEG model as
obtained, e.g., from Monte Carlo simulations.
In Ref.~\cite{Zandi} the amplitude $A_m =  3 \pi^4/(2 u_0)$ (see the text after
Eq.~(\ref{eq:MFF0})) has been estimated beyond MFT
by replacing $u_0$ by  the fixed-point value $u^*$ calculated within
field theory. Although this approach provides a theoretical estimate
for $(A_m)_{theo}= 6.92$, it fails in accounting quantitatively 
for the actual amplitude $(A_m)_{exp}= 1.30\pm 0.03$ observed in experiments
\cite{garcia:06:0}.

For a given film thickness $L$, the position of the minimum of the scaling function corresponds to the
reduced critical temperature
$\tau_m(L)=(T_c(L)-T_{\lambda})/T_{\lambda}= y_m (\xi_0^+/L)^{1/\nu}$  
which reflects the onset temperature $T_c(L)<T_c(L=\infty)=T_{\lambda}$
for superfluidity in  the slab. 
For $\tau>\tau_m$  the
 superfluid  OP profile vanishes and so does
 the mean-field  free energy of the film. Thus from Eqs.~(\ref{eq:1a}) and (\ref{eq:1a1})
it follows that for $T>T_c(L)$ 
one has 
$L^df_C= - L^df_b/(k_BT_{\lambda})\sim - L^d\tau^{2-\alpha}= - (\tau L^{1/\nu})^{d\nu}$, using 
the hyperscaling relation $2-\alpha=d\nu$. For $d=4$ and $\nu=1/2$
this implies  that $\vartheta_0(y_m< y < 0) \sim y^2$ (for $y>0$, within MFT
$f_b=0$ and therefore $f_C=0$) which agrees 
with Eq.~(\ref{eq:MFF1}).

\section{Discussion of the results obtained from the  VBEG model}
\label{sec:5}

\subsection{$^3$He-$^4$He mixtures}
\label{sec:5:mix}

As one can infer  from the comparison of  Figs.~\ref{fig:3} and \ref{fig:3a} 
 the qualitative features
 of  the scaling functions  $\vartheta$ for  $^3$He-$^4$He mixtures
extracted from the experimental data for $X\simeq X_t$, such as 
 the sign of the force, the occurrence of the pronounced 
maximum below $T_t$,  and  the formation of  shoulders above $T_t$,
are   well captured by the present  lattice model. 
The breaks in  slopes 
upon  crossing the $\lambda$-line shown in Fig.~\ref{fig:3}
are features of the mean-field  approach
and  expected to be  smeared out 
by fluctuations.

The experimental data for the Casimir force $f_{C}$  exhibit a maximum
at $tL\simeq -18$\AA \  which cannot be related to the  condition $\xi \sim \zeta \sim L$ 
 borne out by the mean-field  analysis
 (Fig.~\ref{fig:3})  because actually, i.e., beyond MFT, $\xi=\infty$
 in the superfluid phase.
 Further studies are  needed to determine 
 what length scale governs the interfacial width of
the superfluid  OP profile in the 'soft mode' phase below $T_t$.
This analysis, which is left to  future research,  has to take
into account   that  
  the actual  width of the interface formed  in the film 
 (see, e.g.,  the case
$t=-0.0633$ in Fig.~\ref{fig:2}), is broadened both by the Goldstone
 modes in the superfluid phase and
   by capillary-wave like fluctuation.

Different from the mean-field  scaling function $\vartheta$ the experimental 
one  does not vanish  
at low temperatures, which is expected to  be due to the aforementioned
 Goldstone modes of the 
broken continuous  symmetry in the superfluid phase and due to helium-specific
\cite{kardar:04}
surface induced fluctuations which both evade the present mean-field analysis. 
A  similar  behavior has been found in wetting experiments for pure 
 $^4$He films near the $\lambda$-line \cite{garcia:99:0}, in which  
 the film thicknesses  above and below the $\lambda$ transition are
 not the same, so that 
the wetting  films are  thinner in the superfluid phase.
For pure $^4$He Zandi et al.  \cite{kardar:04} pointed out  that the
Goldstone  modes indeed lead to thinner superfluid films for $T \ll T_c$.
But this estimate is not applicable
for  $T\approx T_\lambda$ and  for $T\ll T_\lambda$ it is 
 too small to account for the experimentally observed  magnitude of the 
thinning.  This view of the effect of the Goldstone modes on $\vartheta$ 
is supported by Monte Carlo simulation data for the XY model with periodic BC
\cite{krechanddantchev}. 
The capillary wavelike surface fluctuations,
which  occur on one of the bounding surface of the superfluid $^4$He wetting
 film, give rise to an additional force 
(similar in form but larger in magnitude)
which may then together explain the experimental observation \cite{kardar:04,dkd}.

For a mixture, however, it is possible that the apparent thickening 
of a wetting film as inferred from capacity measurements might be,
at least partially,  an artifact due to  a significant
 change of  the permittivity  within
 the  film~\cite{privat}. Upon inferring  the
 film thickness from the permittivity,  in Ref.~\cite{garcia:02:0} it was
assumed that $X_{film}=X_t$ which does  not hold  at low temperatures at 
 which
the 'soft mode' occurs.
In order to estimate the  error the assumption $X_{film}=X_t$ 
introduces into  the determination of the film thickness
 $L$  we repeat  the calculation for determining  $L$ by  taking
into account  the interface-like {\it concentration}  profile below $T_t$
(see Fig.~\ref{fig:2}) and by assuming a mean field-like shape:
\begin{equation}
\label{eq:MFprof}
X(z)=\frac{1}{2}(X^I+X^{II})-\frac{1}{2}(X^I-X^{II})\tanh[(z-z_0)/(2\zeta)],
\end{equation}
where $X^{I}$ and $X^{II}$ are the  concentrations of the coexisting bulk
phases (see the triangle in Fig.~\ref{fig:1}),  $z_0=L/2$ is the 
position of the center  of the interface, and $\zeta$ is the correlation length
associated with concentration fluctuations. We note that $\zeta$ is 
finite in the
superfluid phase whereas $\xi=\infty$ for the superfluid OP. 
The effective  permittivity constant ${\bar \epsilon}_{film}$ 
of the film follows from adding in series the 
 capacitance  $C$ for each slice  of the film and from using
  $C\sim \epsilon$ \cite{garcia:02:0}:
\begin{equation}
\label{eq:epsilon}
{\bar \epsilon}_{film}(X,T)=\frac{L}{\int_0^{L}{\rm d}z/\epsilon(z)}
\end{equation}
where $\epsilon(z)$ is related to the concentration profile
 via \cite{Kierstead}
\begin{equation}
\label{eq:epsilonconc}
\epsilon(z)-1 = (5.697 - 1.402 X(z))\times 10^{-2}.
\end{equation}
From this we have found that  neglecting at low temperatures
the variation of the 
concentration across the film introduces an error in the determination
of its  thickness from capacity measurements (leading indeed  to 
an increased   film thickness)  which is about 35\% of the
 40\AA\  difference in thickness reported 
above and below $T_t$. Specifically, at  $T=0.65K$ the bulk concentrations are
  $X^I=0.325$,
$X^{II}=0.825$, and the bulk correlation length 
is $\zeta=\zeta_0|t|^{-1}\approx 5.1$\AA, where following 
Ref.~\cite{garcia:02:0}  we have assumed $\zeta_0=1.3$\AA\ as  the
value measured for concentration fluctuations far below $T_t$ in the superfluid
phase. Accordingly,  approximating the actual inhomogeneous permittivity by the homogeneous 
one gives rise to an  error $\approx 14$\AA.

In the crossover regime along the line of critical points connecting the tricritical point and the critical $\lambda$-transition in pure $^4$He 
 only few experimental 
 data for the thicknesses of the wetting 
films  are published. Nonetheless, the observed
  variations of film thicknesses there again agree  with
 the present theoretical
 findings  for the Casimir force.
In particular, one observes a rapid thickening of the films
upon  approaching the line of bulk critical points; 
for specific values of  $X$ a small
 maximum 
located slightly below   the line of bulk critical points
 is also visible (compare Fig.~\ref{fig:3}).

Two reasons impede a more
  quantitative comparison of our results obtained within the
VBEG model with the experimental ones.
First, for our  choice of surface terms in the  Hamiltonian 
the  fixed-point BC $(+,O)$ for the order parameter
 cannot be realized  within the VBEG model.
Taking the limits  $\Delta_1 \to -\infty$ and $\Delta _2\to \infty$
 in  Eqs.~(\ref{eq:set1}) and (\ref{eq:set2}) assures that 
$X(1)=0$  and  $X({\bar L})=1$. 
However, even this limiting concentration profile does not
induce the required BC: although  $m({\bar L})=0$ one has
  $m(1)=I_1(\beta Jb_1)/I_0(\beta Jb_1)\ne 1$, i.e.,
the superfluid  OP at the solid substrate is never saturated at 
its  maximum value 1 which  corresponds to  the BC $(+)$ 
(see Fig.~\ref{fig:profopp}). We have checked
that in this limiting case with respect to $\Delta_{1,2}$
the qualitative behavior of the Casimir force is the same; 
 only the magnitude of $f_C$ is slightly bigger
 ($\vartheta (0)\approx 0.5$ for the limiting case, 
whereas $\vartheta (0)\approx 0.4$ for the case shown in  Fig.~\ref{fig:3}).
 In order to be able to extract universal
properties -- which requires to reach the fixed-point BC -- 
 it would be necessary to introduce a 
 surface field which couples directly 
 to the  superfluid OP so that the BC (+) 
 can be realized; but such a surface field has no physical basis.
 Finally, even at the upper critical dimension
$d=d^{\ast}=3$ due to logarithmic corrections our present 
 MFT is not sufficient. 
However, a naive correction of  $\vartheta$ obtained within the VBEG model
by multiplying it by 
the logarithmic 
factor  $(\ln(L/l_0))^{1/2}$ (see Eq.~(\ref{scaltriC1})) derived within the LG  model does not capture
the proper universal scaling behavior. Instead  renormalization group schemes  
for the VBEG model  have to be employed.

Nonetheless,  our MFT results for the scaling function $\vartheta$ 
within the VBEG model and  for $X=X_t$, 
if  matched with respect to its amplitude with the experimental data
at the tricritical point $y=0$
and after adjusting the scaling variable $y$ by a factor  $y_{th}$
 such  that the experimental 
and theoretical positions
 of the maximum
of the scaling function are  the same (which  is achieved for 
 $y_{th}\approx 0.065$),  yield an adjusted scaling function
${\bar \vartheta} (y)$ which 
reproduces rather  well  the experimental data (see Fig.~\ref{fig:4}),
 especially near the maximum where  interfacial
effects are expected  to be dominant. This observation 
is consistent with  our interpretation that the formation of this maximum
is dominated by the occurrence of  the 
'soft mode' phase which does not 
 depend on the details of the surface fields.  
We note that according to Fig.~\ref{fig:4} 
 the experimental data nominally 
 for $X=X_t$ more closely agree with the theoretical ones  for
$X=X_t-0.01$. This raises the question as to whether
the experimental $^3$He concentration in the film is actually 
 shifted relative 
to the bulk one.

\subsection{Pure $^4$He}
\label{sec:5:pure}

The theoretical models discussed in the previous sections
(VBEG  and LG as lattice and continuum models, respectively) capture 
the {\it
universal} features of  the collective
behavior close to critical (and tricritical) points, such as the Casimir
force. (These models have no predictive power concerning non-universal properties.) The associated finite-size scaling functions acquire  universal forms
if expressed in terms of proper scaling variables, such as 
$L/\xi(\tau)$, where $\xi(\tau)$ is the correlation length 
which controls the large-distance exponential decay of the
two-point correlation functions  of the OP fluctuations {\it in the bulk} at the reduced
temperature $\tau$. 
In systems with discrete symmetry  one has 
$\xi(\tau\rightarrow 0^-) =
\xi_0^-(-\tau)^{-\nu}$ and $\xi(\tau\rightarrow 0^+) =
\xi_0^+\tau^{-\nu}$, where $\xi_0^\pm$ are {\it non-}universal, i.e.,
system-dependent, amplitudes
such that the ratio
$\xi_0^+/\xi_0^-$ is {\it universal} (see, e.g.,
Ref.~\cite{PV}). Accordingly, the scaling function maintains its
universal character also as a function of $y=\tau (L/\xi_0^+)^{1/\nu}$
 in the notation of Sec.~\ref{sec:4}  or, alternatively,  $\tau
(L/\xi_0^-)^{1/\nu}$. 
However, in the case of pure $^4$He, the bulk correlation length
$\xi(\tau<0)$ below the $\lambda$-transition is infinite due to
Goldstone modes and therefore $\xi_0^-$ cannot be defined directly
from the behavior of $\xi(\tau<0)$. Alternatively, one might define
a different length scale $\xi^T(\tau<0) = \xi^T_0(-\tau)^{-\nu}$ 
associated with the power-law decay of
tranverse correlations in the superfluid phase, which is related
to the superfluid density;  the non-universal amplitude $ \xi^T_0$
 forms a universal ratio with $\xi_0^+$ (see, e.g.,
Refs.~\cite{HAHS-76,PV}). For pure $^4$He, experimental estimates of 
$(\xi^T_0)_{exp}$ range from 1.2\AA~\cite{IP-74} to
3.6\AA~\cite{SA-84}, depending on the way it is measured. 
In view of this experimental uncertainty and of the complication
related  to the introduction of $\xi^T_0 \propto \sqrt{u_0}
\xi^+_0$~\cite{HAHS-76} within the MFT discussed in Sec.~\ref{sec:4},
we present the comparison between experimental data and the VBEG model
in terms of the scaling variable $y$, which involves the non-universal
amplitude $\xi_0^+$ the value of which is well assessed experimentally for
$^4$He, $(\xi_0^+)_{exp}=1.43$\AA\ at saturated vapor 
pressure~\cite{ahlers},  and theoretically for the VBEG
model,  $\xi_0^+ = 0.41 a$ within the present MFT, where $a$ is the lattice
spacing (see the end of Subsec.~\ref{subsec:4-1}). Within the LG model one has
an analytic  expression for $\xi_0^+$ in terms of the parameters of the model
(see Eq.~(6.4) in Ref.~\cite{krech:92:a} for  $\xi_0^+$ obtained
within the dimensional regularization scheme).

In Fig.~\ref{fig:fit} we compare  the scaling function obtained 
from the  experimental data for 
the case of pure $^4$He~\cite{garcia:99:0} 
(for a film thickness $L= 423$\AA~\cite{privat})
 with the  MF scaling function $\vartheta_0(y)$ of the  VBEG model
which is  universal for sufficiently thick films. 
The scaling functions are  normalized  by their absolute  values
$|\vartheta_{min}|$ at the  minimum. 
In order to summarize 
all 
available theoretical results we report in the right inset
 of Fig.~\ref{fig:fit} the comparison between
the experimental data for
$T>T_{\lambda}$ and the
scaling function obtained from  
the field-theoretical $\epsilon$-expansion ($\epsilon = 4 -
d$)~\cite{krech:92:a} as follows:
The scaling function $\Theta_{+O,O}(y_+)$ of the 
finite-size contribution  of the renormalized free energy
$f$ provided in Eq.~(6.12) of Ref.~\cite{krech:92:a} has been re-expressed
for $N=2$ (XY model) as a function of $y = \tau (L/\xi_0^+)^{1/\nu}$ via 
$y_+ = y^{1/2}(1 + \epsilon/10 \ln y) + O(\epsilon^2)$ (where $y_+$ is
defined after Eq.~(4.6) in Ref.~\cite{krech:92:a}). 
The resulting expression $\Theta_{+O,O}(y) =
\theta_0(y) + \epsilon\, \theta_1(y) + O(\epsilon^2)$  is then
extrapolated to three dimensions $\epsilon = 1$ either as 
$\Theta_{+O,O}^{\rm [1,0]}(y) = \theta_0(y) + \epsilon\, \theta_1(y)$
(yielding  the solid line in the inset) or 
$\Theta_{+O,O}^{\rm [0,1]}(y) = \theta_0(y)/[1 - \epsilon\,
\theta_1(y)/\theta_0(y)]$ (dashed line), 
corresponding to the Pad\'e approximants
[1,0] and [0,1]. The scaling function of the Casimir force is then
provided by $\vartheta(y) = (d-1) \Theta_{+,O,O}(y) - (y/\nu)
\Theta'_{+,O,O}(y)$ where $d=3$ and $\nu\simeq 0.67$ (see, e.g., 
Table 19 in Ref.~\cite{PV}), accounting for
the actual expression of the scaling variable $y$ in three dimensions.

Discrepancies, such as the position $y_m$ of the minimum, the shape of
the scaling function  for $y>y_m$, the behavior for $y\to -\infty$, and  
the  nonvanishing of $\vartheta ^{exp}$ 
for  $y\ge 0$ can be attributed to  fluctuation effects
neglected in the present MF approach. 
Field-theoretic renormalization group calculations beyond MFT 
yield a quantitative agreement with the experimental
data for $y\ge 0$ \cite{krech:91,krech:92:a,krech:92:b}  
(see Fig.~\ref{fig:fit});
however, so far this field-theoretical approach cannot be extended to
the case $y<0$ \cite{dkd}. From the analysis of Subsec.~\ref{subsec:4-2}
it follows that for fixed $L$ the position
$y_m=-\pi^2\simeq -9.87$ of the minimum is  associated with  the  critical
temperature $T_c(L)$ of the film. 
The experimental data exhibit the position of the minimum at
$x_{min} = -9.8 \pm 0.8 \mbox{\AA}^{1/\nu}$, where 
$x\equiv \tau L^{1/\nu}$~\cite{garcia:99:0,garcia:06:0}, corresponding to
$(y_m)_{exp} \equiv x_{min}/(\xi_0^+)_{exp}^{1/\nu} \simeq -5.7\pm
0.5$  which 
is consistent  with the experimental indication in the sense 
that the onset of superfluidity in the films occurs within the range
$-12 \mbox{\AA}^{1/\nu} \lesssim x \lesssim -7
\mbox{\AA}^{1/\nu}$~\cite{garcia:06:0}, i.e.,  $-8 \lesssim y
\lesssim -5$. But these values of $y$ are  considerably larger than the value
$-\pi^2$ predicted by the  LG approximation. 
In spite of the 
shortcomings mentioned above
the comparison between the experimental and theoretical scaling function is
nonetheless encouraging. The present MF approach does not address the issue 
 that $|\vartheta_{min}/\vartheta(0)|_{exp}\simeq 20$
\cite{garcia:02:0,garcia:06:0} whereas theoretically this ratio is
$\simeq 1$  for periodic BC \cite{dkd}; it is  difficult to expect that
 this ratio
reaches the experimental value  20 corresponding to the actual $(O,O)$ BC.

In passing we mention that in Ref.~\cite{Zandi}
the comparison between Eq.~(\ref{eq:MFF1}) and the experimental data of
Refs.~\cite{garcia:99:0,garcia:06:0} is seemingly affected by an
inconsistent normalization of the experimental and theoretical 
scaling functions which are actually plotted as a function of $\tau
(L/\xi_0^T)^{1/\nu}$ (with $\xi_0^T$ taken from Ref.~\cite{IP-74}) and
$\tau(L/\xi_0^+)^{1/\nu}$, respectively. This artificially 
 reduces the resulting
discrepancy between the experimental and theoretical results 
in comparison to the one displayed in
Fig.~\ref{fig:fit}. 

\section{Summary and Outlook}
\label{sec:6}

\begin{figure}[b]
\includegraphics*[scale=0.6]{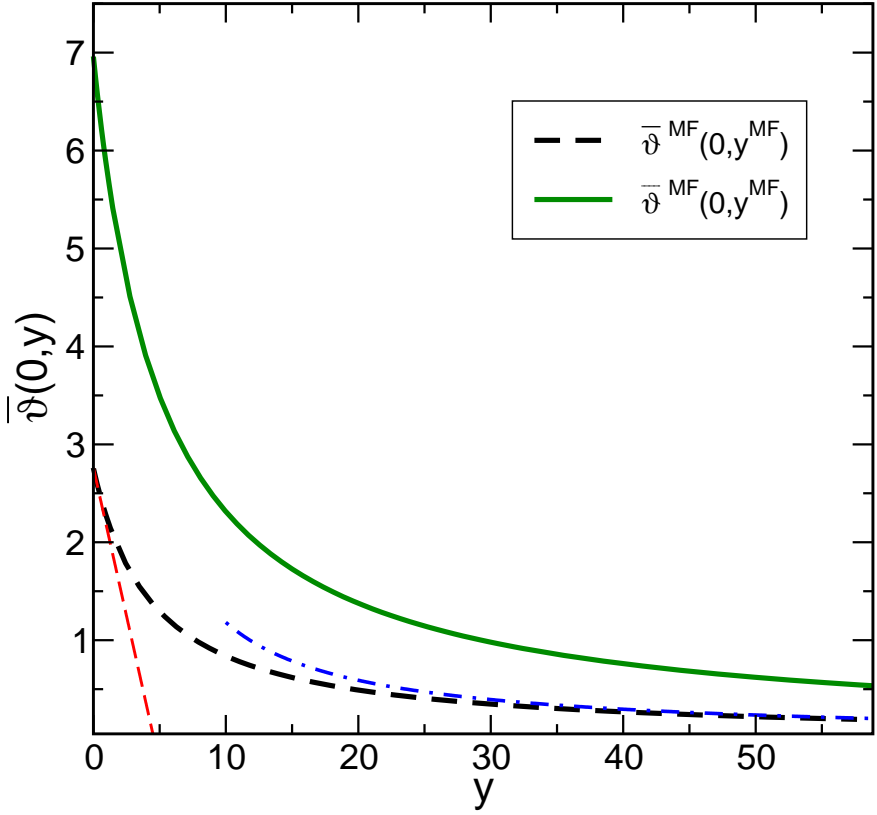}
\caption{Dimensionless MF scaling function 
${\bar \vartheta}^{MF} (r_0L^2=0,y^{MF})=f_CL^3(v_0/90)^{1/2}$
 (see Eq.~(\ref{scaltriC})) 
with  $y^{MF}=(5/(2v_0))^{1/2}u_0L\sim tL$ plotted
 together with the
renormalized mean field  scaling function 
$f_CL^3={\bar \vartheta}^{RMF}(0,y^{RMF})$ 
(see Eq.~(\ref{scaltriC1}) and the  main text) with  $y^{RMF}={\hat u}L(\ln (L/l_0))^{1/14}$, ${\hat u}=7u/(24\pi^2)$, and $L/l_0=400$.
${\bar \vartheta}^{MF}(0,y^{MF}\to\infty)\simeq 11.82/y^{MF}$
(thin dash-dotted line)
and 
 ${\bar \vartheta}^{MF}(0,y^{MF}\to 0)\simeq 2.76-0.605y^{MF}$ (thin dashed line). The asymptotic behavior of ${\bar \vartheta}^{RMF}(0,y^{RMF})$ can 
be obtained from the  one  of ${\bar \vartheta}^{MF}(0,y^{MF})$
 by multiplying the ordinate 
 by the factor $(28/(8\pi^2/3))^{1/2}(\ln (L/l_0))^{1/2}$
and the abscissa by the  factor  $(\ln (L/l_0))^{1/14}$.
 These limiting behaviors have been inferred from asymptotic
expansions of Eq.~(\ref{implicit1}).  }
\label{fig:0}
\end{figure}

\begin{figure}[b]
\includegraphics*[scale=0.6]{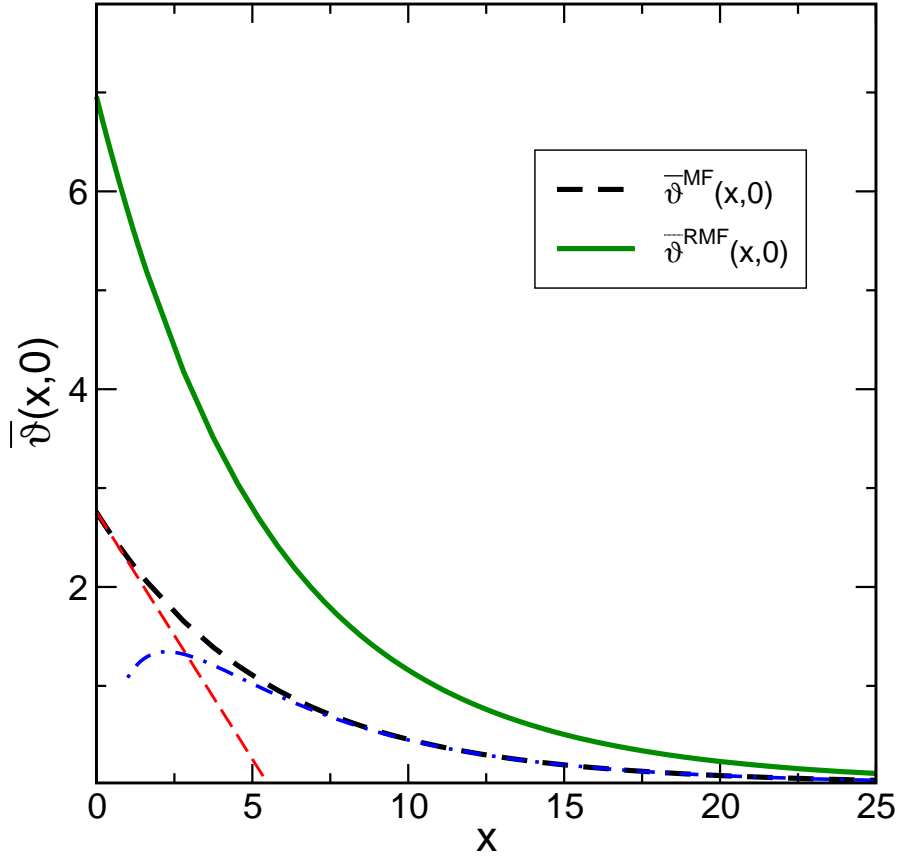}
\caption{Dimensionless MF scaling function 
${\bar \vartheta}^{MF} (x^{MF},u_0=0)=f_CL^3(v_0/90)^{1/2}$
 (see Eq.~(\ref{scaltriC})) 
with  $x^{MF}=x=r_0L^2$ plotted
 together with the
renormalized mean field  scaling function 
$f_CL^3={\bar \vartheta}^{RMF}(x^{RMF},0)$ 
(see Eq.~(\ref{scaltriC1}) and the main text) with  $x^{RMF}=x=rL^2$ 
 and $L/l_0=400$. ${\bar \vartheta}^{MF}(x^{MF}\to\infty,0)\simeq 8(x^{MF})^{3/2}e^{-2(x^{MF})^{1/2}}$  (thin dash-dotted line) and 
 ${\bar \vartheta}^{MF}(x^{MF}\to 0,0)\simeq 2.76-0.5x^{MF}$
(thin dashed line). The asymptotic behavior 
of ${\bar \vartheta}^{RMF} (x^{RMF},0)$ can be obtained from the 
one  of ${\bar \vartheta}^{MF}(x^{MF},0)$
 by multiplying the ordinate  by the factor $(28/(8\pi^2/3))^{1/2}(\ln (L/l_0))^{1/2}$; the abscissa remains the same.
}
\label{fig:0a}
\end{figure}

\begin{figure}[b]
\includegraphics*[scale=0.8]{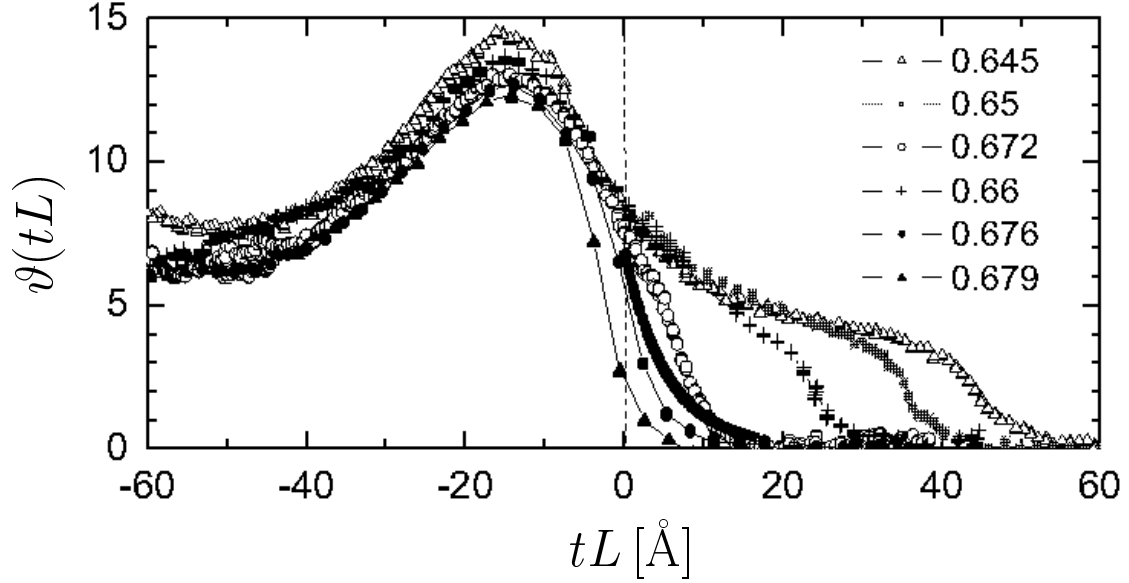}
\caption{Experimental data 
 from Ref.~\cite{garcia:02:0} for the scaling functions
$\vartheta=f_CL^3$ for the Casimir force in $^3$He-$^4$He films
of  thicknesses $L$
along various  paths of fixed $^3$He concentration  (given in the figure)
close to the tricritical concentration $X_t= 0.672$. The scaling variable is in units of \AA.
The solid line
corresponds to  the tricritical  mean-field  scaling function
\cite{garcia:02:0} calculated for $r_0=0$ (i.e., $a=0$ in Eq.~(\ref{ab}))
and suitably adjusted (see the main text); $t=(T-T_t)/T_t$.
\label{fig:3a}}
\end{figure}

\begin{figure}[b]
\includegraphics*[scale=0.6]{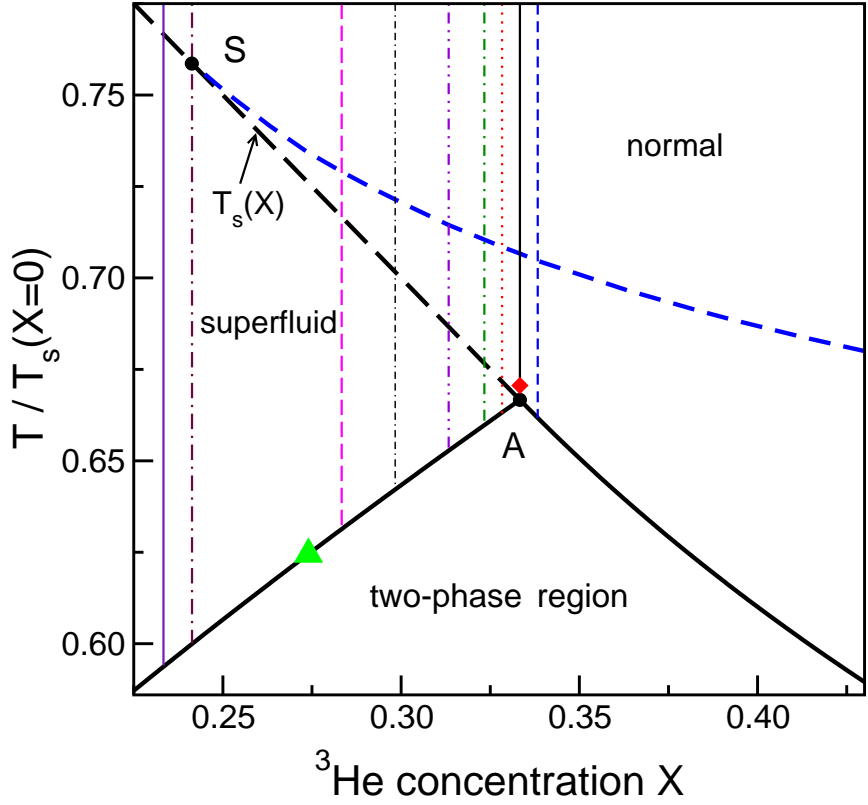}
\caption{ Bulk phase diagram for the VBEG model 
obtained whithin MFT
  for $K/J=0.5$ and $\Delta ^{(l)}/J=-3$ exhibiting the
line $T_s(X)$ of continuous 
 superfluid transitions in the bulk (long-dashed line),
 the  phase separation curves (solid lines), 
the tricritical point  $A=(T_t/T_s(0)=2/3,X_t=1/3)$.
 In a semi-infinite
system there is a  (short-dashed) line of continuous surface transitions
 which merges with the line $T_s(X)$ of bulk critical points
 at the special transition point $S=(T_S/T_s(0)\simeq 0.759, X_S\simeq 0.241$). 
 Upon crossing this surface transition line a thin film near the surface
becomes superfluid although the bulk remains  a normal fluid.
Vertical lines represent thermodynamic paths along which
the Casimir force  has been calculated (see, c.f.,  Fig.~\ref{fig:3}). $\blacklozenge$, $\bullet$ (A) $\blacktriangle$: state points which will be
 considered in Fig.~\ref{fig:2}.\label{fig:1}}
\end{figure}

Based on mean-field analyses of the vectoralized Blume-Emery-Griffiths model
and of the continuum Landau-Ginzburg theory as well as by applying
renormalization group analyses we have obtained the following main results:

(1) By using mean-field theory, near the tricritical point (Fig.~\ref{fig:1})
 we have calculated 
the scaling functions of the Casimir force within  the continuum 
Landau-Ginzburg   theory (Eq.~(\ref{eq:1}))
for the  $O(2)$ model  of  $^3$He-$^4$He  films of thickness $L$
 (see Figs.~\ref{fig:0}  and \ref{fig:0a}).
The scaling functions depend on   two relevant scaling variables
$u_0$ and $r_0$ (see Eq.~(\ref{expan1})).
 By fitting the amplitude of the scaling variable and 
the amplitude of the Casimir force, which remains undetermined
 within the LG mean-field approach, one finds 
  a reasonable agreement with the experimental data
along the thermodynamic path of constant  tricritical  concentration
 of $^3$He (see Fig.~\ref{fig:3a}).

(2) The application of fieldtheoretic   renormalization group analysis
 in spatial dimension $d=3$ yields the correct
 asymptotic  leading behavior of the
Casimir force at the tricritical point. As a function
of the  film thickness   $L$  it has the  form of a power law $\sim L^{-3}$
multiplied by the square root of the logarithm of $L$ and by the universal Casimir
amplitude  (Eq.~(\ref{eq:as})).

(3) Using the fieldtheoretic renormalization group analysis we have derived
the form of the finite-size scaling for the Casimir force 
in the vicinity of the tricritical point and have obtained 
 renormalized mean field  scaling functions
 (see Figs.~\ref{fig:0} and \ref{fig:0a}). It turns out that
 also one of the scaling variables
 acquires a logarithmic correction (Eq.~(\ref{scaltriC1})).

\begin{figure}[b]
\includegraphics*[scale=0.6]{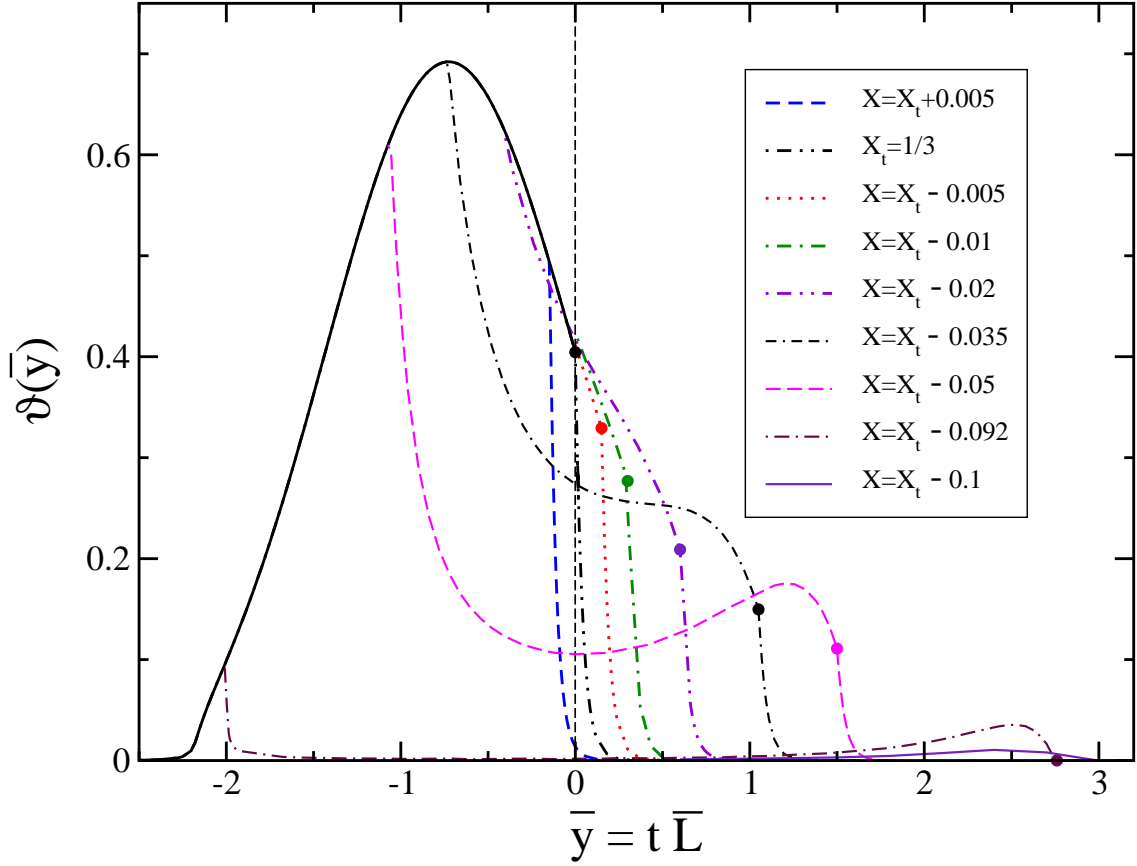}
\caption{Dimensionless scaling function 
${\vartheta}({\bar y}=t{\bar L})=f_C{\bar L}^3$, with $t=(T-T_t)/T_t$ and
${\bar L}=20$
 for the Casimir force calculated within MFT for the VBEG model
along the paths of fixed  concentration of $^3$He shown in Fig.~\ref{fig:1}.
Dots indicate the corresponding  onset temperature $T_s(X)$ of  superfluidity
at the line of bulk critical points. The full line for ${\bar y}<0$ corresponds to the temperatures of the onset of the first-order phase separation in the bulk (see Fig.~\ref{fig:1}). In view of, c.f., Fig.~\ref{fig:limsc} we note that the curves might still shift if calculated for larger values of ${\bar L}$. 
\label{fig:3}}
\end{figure}

\begin{figure}[b]
\includegraphics*[scale=0.6]{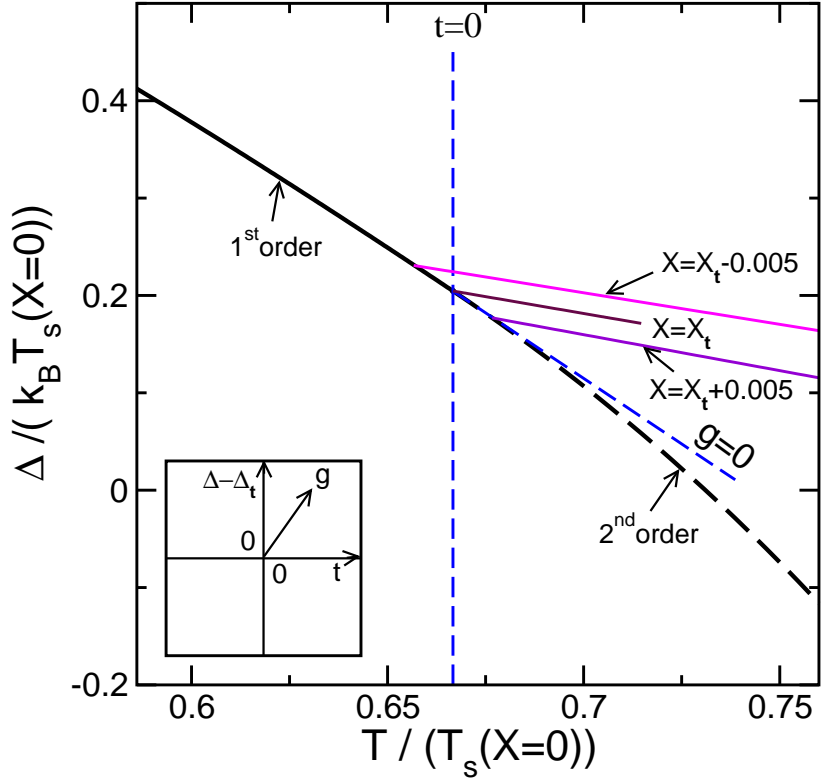}
\caption{ Bulk phase diagram for the VBEG model in the $(\Delta, T)$ plane
obtained whithin MFT for the same set of parameters as in Fig.~\ref{fig:1}. 
The long-dashed coexistence line corresponds to the  continuous 
 superfluid transitions  whereas the solid coexistence line corresponds to
 the curve of first-order phase separation. 
As indicated in the inset 
$g$ and $t$ are the 
two relevant scaling variables  (compare Eq.~(\ref{scalfields}));
the line  $g=0$ is  tangential  to the coexistence line 
at the tricritical point where the lines of first-  and second-order
transitions merge. 
Note that according to Eq.~(\ref{scalfields}) along the line $g=0$
 one has $(\Delta-\Delta_t)/(k_BT_t)=-a't$ and along the line $t=0$ one has
 $g=(\Delta-\Delta_t)/(k_BT_t)$.
Three thermodynamic  paths of constant concentration are shown:
$X=X_t$, $X=X_t-0.005$ (upper line), and $X=X_t+0.005$ (lower line).
We note
that along the paths of constant concentration both scaling variable $t$ and $g$ vary; however, the variation of $t$ is more pronounced so that within a rough
approximation $g$ can be considered to be constant along each path.
\label{fig:1a}}
\end{figure}

\begin{figure}[b,h]
\includegraphics*[scale=0.6]{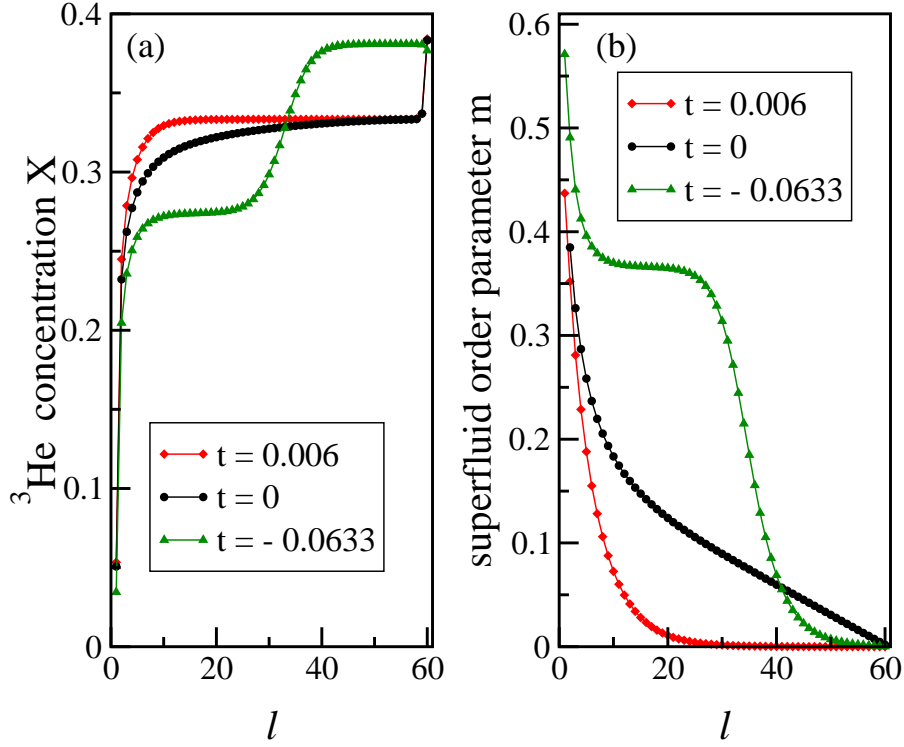}
\caption{(a) $^3$He concentration profile $X(l)=1-Q_l$   and (b) superfluid 
OP profile $m_l$ for a VBEG film of thickness
${\bar L}=60$ for  $K=0.5J$, 
$\Delta ^{(l)}/ J=-3$, and $\Delta^{(r)}/ J=\Delta_t/ J\simeq  0.61$
corresponding to the state points $\blacklozenge$, $\bullet$, and 
$ \blacktriangle$ indicated in Fig.~\ref{fig:1}; $t=(T-T_t)/T_t$.
 \label{fig:2}}
\end{figure}

\begin{figure}[b,h]
\includegraphics*[scale=0.6]{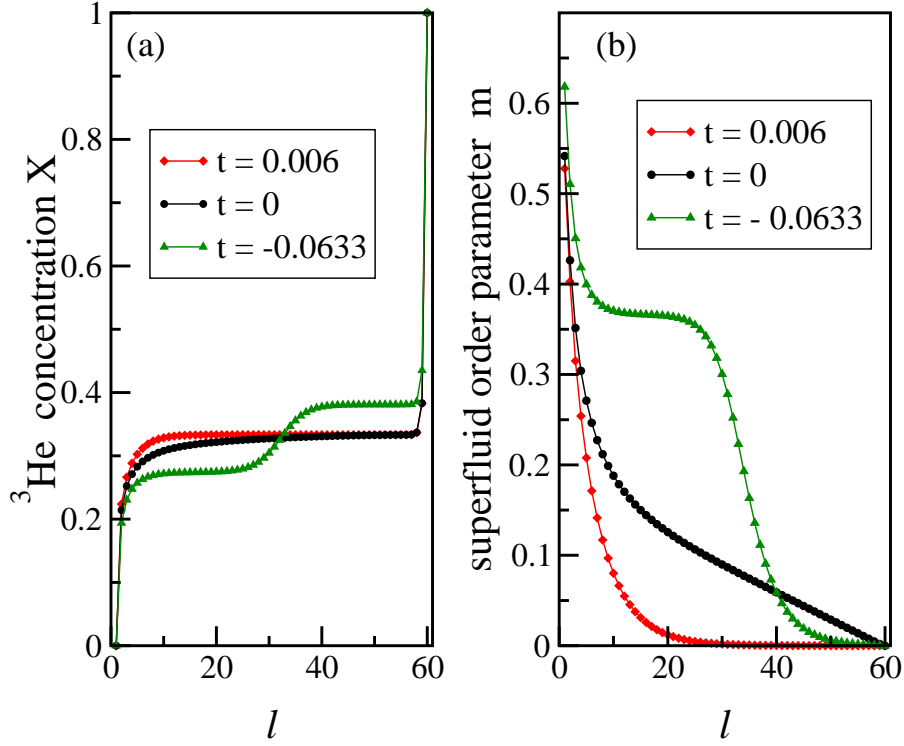}
\caption{(a) $^3$He concentration profile $X(l)=1-Q_l$   and (b) superfluid 
OP profile $m_l$ for a VBEG film of  width
${\bar L}=60$ for  $K=0.5J$, 
$\Delta ^{(l)}/ J=-\infty$, and $\Delta^{(r)}/ J=+\infty$
corresponding to the state points $\blacklozenge$, $\bullet$, and 
$ \blacktriangle$ indicated in Fig.~\ref{fig:1} with $t=(T-T_t)/T_t$.}
 \label{fig:profopp}
\end{figure}

(4) Using  mean-field approximation 
we have calculated the scaling function of the Casimir force
within the vectoralized Blume-Emery-Griffith
lattice model  of  $^3$He-$^4$He mixtures  along the thermodynamic
 paths of 
fixed $^3$He concentrations  (see Figs.~\ref{fig:1}, \ref{fig:1a}, and \ref{fig:3}). For concentrations
of $^3$He close to the tricritical concentration our results are in a 
qualitative agreement with the available 
 experimental data (see Figs.~\ref{fig:3a}, \ref{fig:3}, and \ref{fig:4}).
Our calculations also predict the crossover behavior of the Casimir
force along the line of critical points connecting the tricritical
point and the $\lambda$-transition  for pure $^4$He.
We have found that the pronounced maximum of the Casimir force, which 
occurs  below the tricritical temperature, is associated with the formation
of a  'soft mode'  phase within the film (see Figs.~\ref{fig:2} 
and \ref{fig:profopp}).

\begin{figure}[b]
\includegraphics*[scale=0.6]{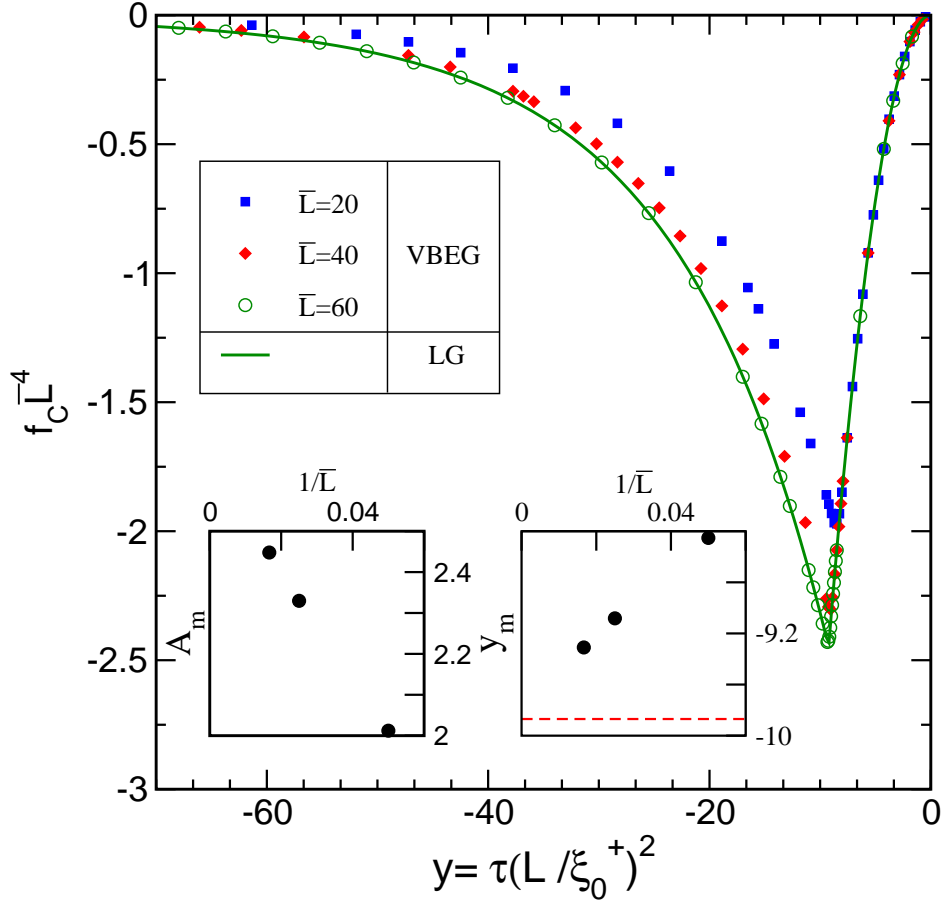}
\caption{Mean-field scaling function 
$\vartheta_0(y=\tau( L/\xi_0^+)^2)=f_C{\bar L}^4$
  for the limiting case of the 
 VBEG model (symbols) corresponding to pure $^4$He and various film thicknesses
${\bar L}$ with  $\tau=(T-T_{\lambda})/T_{\lambda}$.
The full curve corresponds to the 
scaling function ${\bar \vartheta}_0^{LG}(y)$ obtained from the  continuum
$O(2)$ LG  theory within MFT (Eqs.~(\ref{eq:MFF3}) and (\ref{eq:MFF1}))
whith the amplitude $A_m = A_m(\bar L)$ and the position of the
minimum $y_m = y_m(\bar L)$ determined  in such a way as  to provide the
best fit to $\vartheta_0$ from the VBEG model;
for further details see the main text. With this rescaling
the continuum theory provides a very good fit (here shown only for $\bar L
= 60$) to the numerical data. 
The insets show the ${\bar L}$-dependence of $A_m$ 
and $y_m$ used as fitting parameters. The dashed line in the inset 
for $y_m({\bar L})$ indicates the limiting value $y_m=-\pi^2$ 
predicted by th LG model.
Surprisingly, 
scaling -- corresponding to $\bar L$-independent $A_m$ and $y_m$ -- is
not yet attained by the numerical data of the VBEG 
model even for thick slabs with ${\bar L}\simeq 60$.
}
\label{fig:limsc}
\end{figure}

\begin{figure}[b,h]
\includegraphics*[scale=0.6]{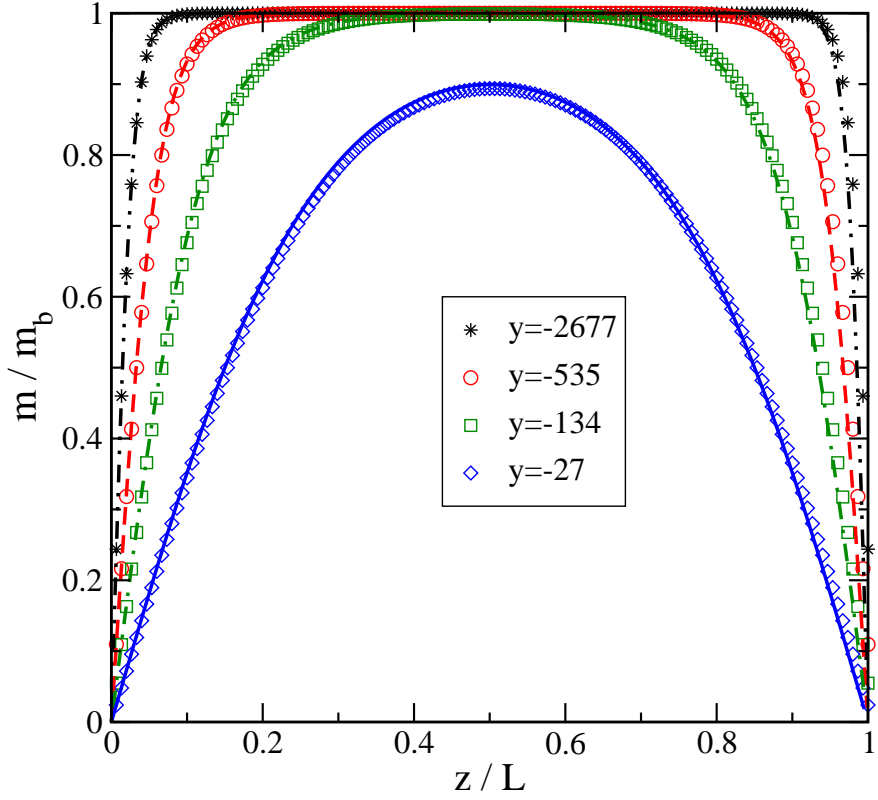}
\caption{Mean-field OP  profiles (normalized to the corresponding bulk values
$m_b$) across slabs of thickness $L$   calculated from the limiting case of the VBEG model (symbols,  $\bar L = 150$) and from
the continuum
$O(2)$ LG theory (lines, see Eqs.~(202) and (203) in
Ref.~\cite{andrea}) for a  selection of the 
scaling variable $y= \tau (L/\xi_0^+)^{1/\nu}$ below the shifted
critical point of the film (corresponding to $y=y_m=-\pi^2$, see the main
text). For $y$ sufficiently negative $m(z\gg a)-m_b \sim \exp(-z/\xi(\tau<0))$
in the middle of the slab. This allows one to infer ${\bar \xi}_0^-={\bar \xi}(\tau<0)(-\tau)^{1/2}\simeq 0.29$ so that ${\bar \xi}_0^+={\sqrt 2}{\bar\xi}_0^-\simeq 0.41$.
}
 \label{fig:9}
\end{figure}

\begin{figure}[b]
\includegraphics*[scale=0.6]{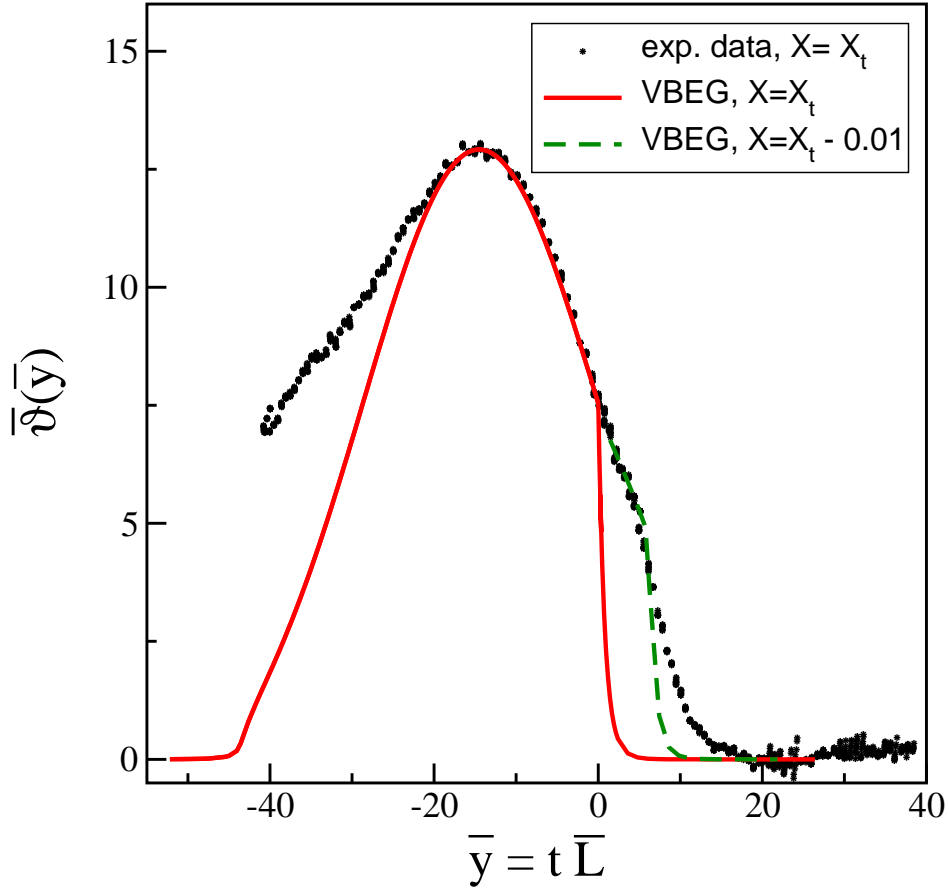}
\caption{The adjusted scaling function ${\bar \vartheta}({\bar y})$ (see the 
main text)
 for the  VBEG model within  MFT compared with the corresponding  
experimental curve \cite{garcia:02:0} obtained along the path of
 fixed tricritical
 concentration $X=X_t\approx 0.672$ of $^3$He. ${\bar \vartheta}({\bar y})$ is obtained
from $ \vartheta({\bar y})$ in Fig.~5 by rescaling the amplitudes of $\vartheta$ 
and ${\bar y}$ such that there is agreement between the experimental data for 
$X=0.672$ at ${\bar y}=0$ and with respect to the positions of the maximum.
The VBEG curve for $X=X_t-0.01$ agrees with the experimental data for nominally
$X=X_t$ even better. Both theoretical curves coincide for ${\bar y}<0$.
\label{fig:4}}
\end{figure}

\begin{figure}[b]
\includegraphics*[scale=0.6]{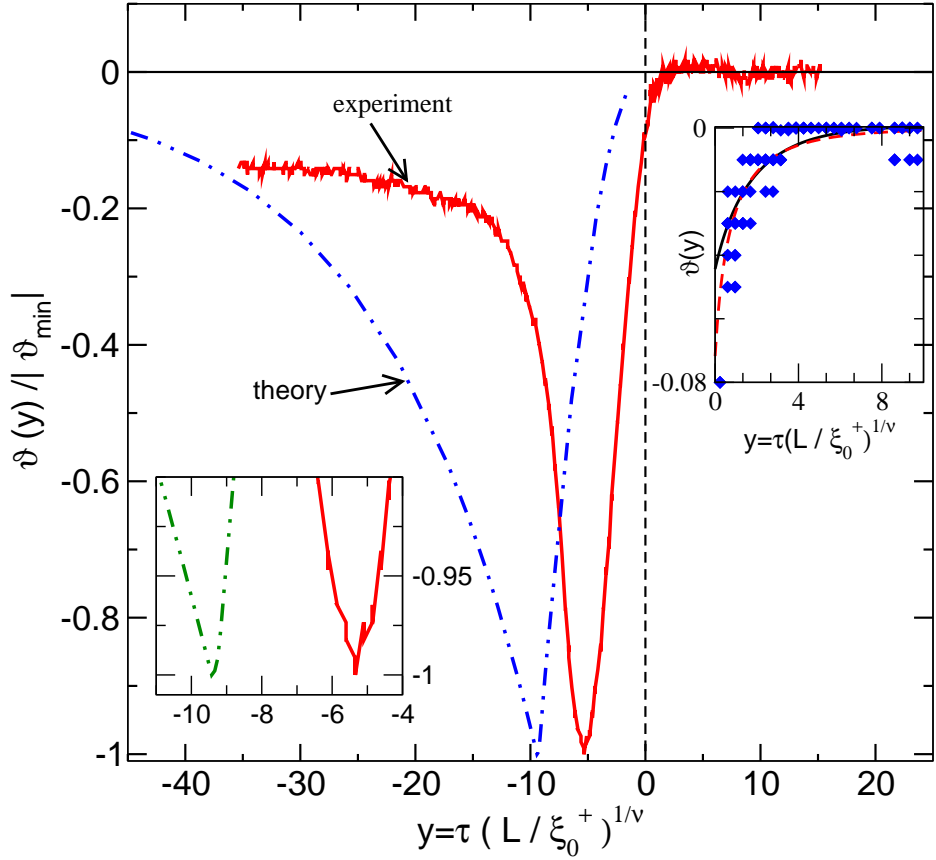}
\caption{%
Normalized  mean-field  scaling function ${\vartheta}_0(y)$ 
for the limiting case of the VBEG model (on a lattice with $\bar L = 60$)
corresponding to  pure $^4$He  compared with the 
experimental data $(\vartheta)_{exp}$ \cite{garcia:99:0} in terms of the
proper scaling variable $y=\tau(L/\xi_0^+)^{1/\nu}$ using
$(\xi_0^+)_{exp}=1.43$\AA\ for pure $^4$He~\cite{ahlers} 
and $\nu=0.67$.
These are the universal forms of the scaling function
$\vartheta_0$. 
The inset on the left shows a magnification of the main plot close to
the minimum.  According to the analysis presented in  
Subsec.~\protect{\ref{subsec:4-2}} (see
also Fig.~\protect{\ref{fig:limsc}}) 
the position $y_m(\bar L)$ 
of the minimum of the theoretical curve in the scaling
limit $\bar L=\infty$ approaches the value $-\pi^2$.
In the inset on the right the experimental data
 (diamonds) above the 
critical temperature
are compared with  the scaling functions for the three-dimensional XY
model in a slab obtained from the $\epsilon$-expansion (see
the main text: the solid (dashed) line corresponds to the [1,0] ([0,1])
Pad\'e approximant). Due to the experimental resolution $(\vartheta)_{exp}$
takes only discretized values.
\label{fig:fit}}
\end{figure}

(5) We have analyzed the limiting case of the VBEG model which corresponds to the classical XY model for pure $^4$He. Within mean-field theory we have been
 able to show that for sufficiently thick films
the   scaling functions as obtained from the lattice model for the Casimir force are in an agreement with the ones obtained from the continuum  $O(2)$
Landau-Ginzburg theory  (see Fig.~\ref{fig:limsc}). The encouraging 
comparison of
 the former with the experimental data is displayed in Fig.~\ref{fig:fit}.

As an outlook we propose to test experimentally  the
scaling of the  Casimir force  for different 
thicknesses of the wetting films by  taking into account
logarithmic corrections. Moreover it appears to be promising
to study experimentally in more detail the crossover
of the Casimir forces  between their  tricritical behavior  and
their critical behavior near the  $\lambda$-transition 
and to compare it with the  theoretical
predictions  presented here.

A.M. benefited from discussions with R. Garcia and M. Krech.

\vfill\eject

\vfill \eject \break

\vfill \eject \break

\vfill \eject \break

\vfill \eject \break

\vfill \eject \break

\vfill \eject \break

\vfill \eject \break

\vfill \eject \break

\vfill \eject \break

\begin{thebibliography}{99}

\bibitem{casimir} H. B. Casimir, Proc. K. Ned. Akad. Wet. {\bf 51}, 793 (1948).


\bibitem{garcia:99:0} R. Garcia and M. H. W. Chan, Phys. Rev. Lett. {\bf 83}, 1187 (1999).


\bibitem{law:99:0}A. Mukhopadhyay and B. M. Law, Phys.~Rev.~Lett. {\bf 83}, 772 (1999).


\bibitem{garcia:02:0} R. Garcia and M. H. W. Chan, Phys. Rev. Lett. {\bf 88}, 086101 (2002).

\bibitem{balibar:02:0} T. Ueno, S. Balibar, T. Mizusaki, F. Caupin, and 
E. Rolley, Phys. Rev. Lett. {\bf 90}, 116102 (2003); R. Ishiguro and S. Balibar, J. Low Temp. Phys. {\bf 140}, 29 (2005).

\bibitem{pershan} M. Fukuto, Y. F. Yano, and P. S. Pershan, Phys. Rev. Lett. {\bf 94}, 135702 (2005).

\bibitem{garcia:06:0} A. Ganshin, S. Scheidemantel, R. Garcia, and M. H. W. Chan, Phys. Rev. Lett. {\bf 97}, 075301 (2006).

\bibitem{fdg} M. E. Fisher and P. G. de Gennes, C. R. Acad. Sci. Paris Ser. B {\bf 287}, 207 (1978).


\bibitem{krech:99:0} M. Krech, {\it The Casimir Effect in  Critical System}  
(World Scientific, Singapore, 1994); J. Phys. Condens. Matter {\bf 11}, R391 (1999);    M. P. Nightingale and J. O. Indekeu, Phys. Rev. Lett. {\bf 54}, 1824 (1985); J. Indekeu, J. Chem. Soc. Faraday Trans. II {\bf 82}, 1838 (1986).

\bibitem{krech:91} M. Krech and S. Dietrich, Phys. Rev. Lett. {\bf 66}, 345 (1991); {\it ibid} {\bf 67}, 1055 (1991).

\bibitem{privman} V. Privman, in {\it Finite Size Scaling and Numerical Simulation of Statistical Systems}, edited by V. Privman (World Scientific, Singapore, 1990), p. 1.


\bibitem{diehl:86:0} H. W. Diehl, in {\em Phase Transitions and Critical
Phenomena}, edited by C. Domb and J. L. Lebowitz (Academic, London,
1986), Vol. 10, p.76.

\bibitem{krech:92:a}  M. Krech and S. Dietrich, Phys. Rev. A {\bf
46}, 1886 (1992). 

\bibitem{krech:92:b}  M. Krech and S. Dietrich, Phys. Rev. A {\bf 46}, 1922 (1992).

\bibitem{kardar:04} R. Zandi, J. Rudnick, and M. Kardar, Phys. Rev. Lett. {\bf 93}, 155302 (2004).

\bibitem{riedel:72:0} E. K. Riedel, Phys. Rev. Lett. {\bf 28}, 675 (1972);
 E. K. Riedel and F. J. Wegner, Phys. Rev. Lett. {\bf 29}, 349 (1972).

\bibitem{LawSar}
D. Lawrie and S. Sarbach, in {\em Phase Transitions and Critical
Phenomena}, edited by C. Domb and J. L. Lebowitz (Academic, London,
1984), Vol. 9, p.2.

\bibitem{laheurte:78:0} J.-P. Romagnan, J.-P. Laheurte, J.-C. Noiray, and W. F. Saam, J. Low Temp. Phys.~{\bf 30}, 425 (1978).

\bibitem{maciolek:06:0} A. Macio\l ek and S. Dietrich, Europhys. Lett. {\bf 74}, 22 (2006). 

\bibitem{maciolek:04:0} A. Macio\l ek, M. Krech, and S. Dietrich, Phys. Rev. E
{\bf 69}, 036117 (2004); and references therein.

\bibitem{eisen:88} E. Eisenriegler and H. W. Diehl, Phys. Rev. B {\bf 37}, 5257 (1988); and references therein.


\bibitem{riedel:72} E. K. Riedel, Phys. Rev. Lett. {\bf 28}, 675 (1972). 


\bibitem{leiderer} P. Leiderer, D. R. Watts, and W. W. Webb, Phys. Rev. Lett. {\bf 33}, 483 (1974).



\bibitem{ritschel} U. Ritschel and M. Gerwinski, Physica A {\bf 243}, 
362 (1997). 

\bibitem{footnote}  For critical systems in  the film geometry the
renormalization of the free energy was discussed  in Ref.~\cite{krech:92:a}.
It was shown that  additive terms give rise to 
 finite-size contributions
to the free energy which are analytic in $t=(T-T_c)/T_c$
and exponentially small as a function of $L$.

\bibitem{amit} D. J. Amit, {\em Field theory, the Renormalization Group and Critical Phenomena} (McGraw Hill, New York, 1978).



\bibitem{bell:89:0}  G. M. Bell and D. A. Lavis, {\it Statistical Mechanics
 of Lattice Models}, series  "Mathematics and its Applications"  (Ellis Horwood Ltd, Chichester, 1989).

\bibitem{book}  P. M. Chaikin and T. C. Lubensky, {\it Principles of
Condensed Matter Physics} (Cambridge University Press, 1995).

\bibitem{peliti:85:0} A. Crisanti and L. Peliti, J. Phys. A: Math. Gen.
 {\bf 18},  L543 (1985).

\bibitem{parry:92:0} A. O. Parry and R. Evans, Phys. Rev. Lett. {\bf 64}, 439 (1990).

\bibitem{stecki}  R. Evans  and J. Stecki, Phys. Rev. B {\bf 49}, 8842 (1994); 
A. O. Parry and R. Evans, Phys. Rev. Lett. {\bf 64}, 439 (1990).

\bibitem{fisher:86:0} M. E. Fisher, J. Chem. Soc. Faraday Trans. II {\bf 82}, 1569 (1986).
 
\bibitem{kosterlitz:80} J. M. Kosterlitz and D. J. Thouless,
J. Phys. C: Solid State Phys. {\bf 6}, 1181  (1973). 

\bibitem{andrea} A. Gambassi and S. Dietrich, J. Stat. Phys. {\bf 123}, 929 
(2006).

\bibitem{fisher} M. E. Fisher and H. Nakanishi, J. Chem. Phys. {\bf 75}, 5857 (1981).

\bibitem{Zandi} 
R. Zandi, A. Shackell, J. Rudnick, M. Kardar, and L. P. Chayes,
preprint cond-mat/0703262.

\bibitem{krechanddantchev} D. Dantchev and M. Krech, Phys. Rev. E {\bf 69}, 046119 (2004).

\bibitem{dkd} D. Dantchev, M. Krech, and S. Dietrich, Phys. Rev. Lett. {\bf 95}, 259701 (2005). 

\bibitem{privat} R. Garcia, private communication. 

\bibitem{Kierstead} H. A. Kierstead, J. Low Temp. Phys. {\bf 24}, 497 (1976).

\bibitem{PV} 
A. Pelissetto and E. Vicari, Phys. Rep. {\bf 368}, 549 (2002).

\bibitem{HAHS-76}
P. C. Hohenberg, A. Aharony, B. I. Halperin, and E. D. Siggia,
Phys. Rev. B {\bf 13}, 2986 (1976).

\bibitem{IP-74}
G. G. Ihas and F. Pobell,
Phys. Rev. A {\bf 9}, 1278 (1974).

\bibitem{SA-84}
A. Singsaas and G. Ahlers,
Phys. Rev. B {\bf 30}, 5103 (1984).

\bibitem{ahlers} W. Y. Tam and G. Ahlers, Phys. Rev. B {\bf 32}, 5932
(1985), Table XI.

\end{thebibliography}
\end{document}